\documentclass[12pt]{article} 
\pdfoutput=1

\usepackage[utf8]{inputenc}
\usepackage[T1]{fontenc} 
\usepackage{upgreek} 
\usepackage{lmodern}
\usepackage[english]{babel}      
\usepackage{cancel}
\usepackage[pdftex]{graphicx}
\usepackage{epsfig}
\usepackage{graphicx}
\usepackage{comment}
\usepackage{latexsym}
\usepackage{hyperref}
\usepackage{amsmath}
\usepackage{physics}
\usepackage{mathrsfs}
\usepackage[usenames, dvipsnames]{color}
\usepackage{amsbsy}
\usepackage{amssymb}
\usepackage{amsthm}
\usepackage{amsfonts}
\usepackage{cite}
\usepackage{enumitem}
\usepackage{xcolor}
\usepackage{color}
\usepackage{mathtools}
\usepackage{caption} 
\usepackage{subcaption} 
\usepackage{euscript} 
\usepackage{float}
\usepackage{diagbox}
\usepackage[title]{appendix}
\usepackage{tabularx}
\usepackage{tikz}
\usetikzlibrary{calc}
\usepackage{pgfplots} 
\pgfplotsset{compat=1.17} 
\usepackage{tcolorbox}
\usetikzlibrary{decorations.pathmorphing,arrows.meta}

\usepackage{amsmath}
\allowdisplaybreaks
\usepackage{amsfonts}
\usepackage{amssymb}
\usepackage{bbm}
\usepackage{verbatim}
\usepackage{dsfont}
\usepackage{booktabs}
\usepackage{slashed}
\usepackage{braket}
\usepackage{soul}

\renewcommand\d{{\rm d}}

\newcommand{\be}{\begin{equation}}
\newcommand{\ee}{\end{equation}}

\renewcommand{\thefootnote}{\fnsymbol{footnote}}
\newcommand{\eq}[1]{equation \eqref{#1}}

\newcommand{\Sect}[1]{Section~\ref{#1}}
\newcommand{\Sects}[1]{Sections~\ref{#1}}
\newcommand{\Appendix}[1]{Appendix~\ref{#1}}
\newcommand{\Fig}[1]{Figure~\ref{#1}}

\newcommand{\Le}{{\rm L}}
\newcommand{\Ri}{{\rm R}}

\newcommand{\ie}{{\it i.e.} }
\newcommand{\eg}{{\it e.g.} }

\newcommand{\apriori}{{\it a priori} }

\renewcommand{\and}{\mbox{and}}

\newcommand{\bm}{\boldmath} 

\catcode`\@=11
\def\marginnote#1{}
\newcount\hour
\newcount\minute
\newtoks\amorpm
\hour=\time\divide\hour by60 \minute=\time{\multiply\hour by60
\global\advance\minute by-\hour}
\edef\standardtime{{\ifnum\hour<12 \global\amorpm={am}%
        \else\global\amorpm={pm}\advance\hour by-12 \fi
        \ifnum\hour=0 \hour=12 \fi
        \number\hour:\ifnum\minute<10 0\fi\number\minute\the\amorpm}}
\edef\militarytime{\number\hour:\ifnum\minute<10 0\fi\number\minute}
\def\draftlabel#1{{\@bsphack\if@filesw {\let\thepage\relax
   \xdef\@gtempa{\write\@auxout{\string
      \newlabel{#1}{{\@currentlabel}{\thepage}}}}}\@gtempa
   \if@nobreak \ifvmode\nobreak\fi\fi\fi\@esphack}
        \gdef\@eqnlabel{#1}}
\def\@eqnlabel{}
\def\@vacuum{}
\def\draftmarginnote#1{\marginpar{\raggedright\scriptsize\tt#1}}
\def\draft{\oddsidemargin -.2truein
        \def\@oddfoot{\sl preliminary draft \hfil
        \rm\thepage\hfil\sl\today\quad\militarytime}
        \let\@evenfoot\@oddfoot \overfullrule 3pt
        \let\label=\draftlabel
        \let\marginnote=\draftmarginnote
   \def\@eqnnum{(\theequation)\rlap{\kern\marginparsep\tt\@eqnlabel}%
\global\let\@eqnlabel\@vacuum}  }
\def\thebibliography#1{
\vskip 0.5cm \centerline{\bf \Large References}
\list{
[\arabic{enumi}]}{\settowidth\labelwidth{[#1]}
\leftmargin\labelwidth
\advance\leftmargin\labelsep
\usecounter{enumi}}
\def\newblock{\hskip .11em plus .33em minus .07em}
\sloppy\clubpenalty4000\widowpenalty4000
\sfcode`\.=1000\relax}

\renewcommand{\theequation}{\arabic{section}.\arabic{equation}}
\renewcommand{\section}{\setcounter{equation}{0}\@startsection
{section}{1}{0mm}{-\baselineskip}{0.5\baselineskip} {\normalfont\Large\bfseries}}
\renewcommand{\subsection}{\@startsection
{subsection}{2}{0mm}{-\baselineskip}{0.5\baselineskip} {\normalfont\large\bfseries}}
\renewcommand{\subsubsection}{\@startsection
{subsubsection}{3}{0mm}{-\baselineskip}{0.5\baselineskip}
{\normalfont\normalsize\slshape}}

\begin{document}


\begin{titlepage}
\begin{flushright}
CPHT-RR074.112023, November 2023
\vspace{0.0cm}
\end{flushright}
\begin{centering}
{\bm\bf \Large On the Quantum Bousso Bound in JT gravity}

\vspace{6mm}

 {\bf Victor Franken,$^1$\footnote{victor.franken@polytechnique.edu} Fran\c{c}ois Rondeau$^2$\footnote{rondeau.francois@ucy.ac.cy}}

 \vspace{3mm}

$^1$  {\em CPHT, CNRS, Ecole polytechnique, IP Paris, \\F-91128 Palaiseau, France}

$^2$ {\em Department of Physics, University of Cyprus, \\Nicosia 1678, Cyprus}

\end{centering}
\vspace{0.5cm}
$~$\\
\centerline{\bf\large Abstract}\vspace{0.2cm}
\vspace{-0.6cm}

\begin{quote}
We prove the Strominger-Thompson quantum Bousso bound in the infinite class of conformal vacua in semiclassical JT gravity, with postive or negative cosmological constant. The Bousso-Fisher-Leichenauer-Wall quantum Bousso bound follows from an analogous derivation, requiring only initial quantum non-expansion. In this process, we show that the quantity ${2\pi k^{\mu}k^{\nu}\braket{:T_{\mu\nu}:}-S''-\frac{6}{c}(S')^2}$ vanishes in any vacuum state, entailing a stronger version of Wall's quantum null energy condition. We derive an entropy formula in the presence of a generic class of two reflecting boundaries, in order to apply our argument to the half reduction model of de Sitter JT gravity.
\end{quote}

\newpage
\thispagestyle{empty}
\tableofcontents
\newpage

\end{titlepage}
\newpage
\setcounter{footnote}{0}
\renewcommand{\thefootnote}{\arabic{footnote}}
 \setlength{\baselineskip}{.7cm} \setlength{\parskip}{.2cm}

\setcounter{section}{0}


\section{Introduction}

In the context of black hole thermodynamics, the famous black hole entropy formula\footnote{Throughout this paper, we will work in natural units, setting $\hbar=1$.}
\begin{equation}
    S_{\rm BH} = \frac{A}{4\hat{G}},
\end{equation}
where $A$ is the area of the black hole horizon and $\hat{G}$ the gravitational Newton's constant, must be generalized for the second law of thermodynamics to hold \cite{Hawking:1971tu,Bekenstein1,Bekenstein2,Bekenstein3}. To this purpose, Bekenstein introduced the notion of generalized entropy
\begin{equation}
\label{eq:sgen}
    S_{\rm gen} = \frac{A}{4\hat{G}} + S_{\rm out},
\end{equation}
where $S_{\rm out}$ is the coarse-grained (or thermodynamic) entropy of matter outside the black hole, which satisfies the generalized second law \cite{Wall:2009wm, Wall:2011hj}
\begin{equation}
    \d S_{\rm gen}\geq 0.
\end{equation}
Building on this connection between spacetime geometry and information (entropy), 't Hooft and Susskind developed the holographic principle \cite{tHooft:1993dmi, Susskind:1994vu}, associating the information contained in a region of spacetime to the area of a lower dimensional surface. This led to the development of the AdS/CFT correspondence \cite{Maldacena:1997re}, relating the gravitational theory in Anti de Sitter spacetime with a conformal field theory living on the boundary of this spacetime. Bousso then formalized the holographic principle by formulating a covariant entropy bound \cite{Bousso:1999xy, Bousso:2002ju}, whose statement goes as follows. Consider a codimension $2$ spacelike surface $B$ and a codimension $1$ null hypersurface $L$ emanating from it. If $L$ is of non-positive expansion, then the coarse-grained entropy $S_{L}$ passing through $L$ is bounded from above by one quarter the area of $B$, in Planck units. Later, the Bousso bound has been generalized to the case where $L$ terminates on a second codimension $2$ surface $B'$ \cite{Flanagan:1999jp}:
\begin{equation}
    S_{L}\leq \frac{1}{4\hat{G}}\Delta A,
\end{equation}
where $\Delta A$ is the difference of the areas of $B$ and $B'$.\footnote{These notions will be defined more carefully in \Sect{sec:lightsheets} and Appendix \ref{sec:hydro_regime}.} Note that in order to provide a formal proof of the Bousso bound, one needs to precisely define the quantity $S_L$, which is ambiguous \apriori. In particular, the concept of entropy ``passing through'' a light-sheet implies a notion of local entropy. As we will see in Section \ref{sec:lightsheets}, such a definition is obtained in the ``hydrodynamic regime'', where the matter coarse-grained entropy (\ie thermodynamic entropy) has a phenomenological description in terms of a local entropy current. In this regime, the classical Bousso bound can be proven rigorously \cite{Flanagan:1999jp, Bousso:2003kb}.

More recently, Ryu and Takayanagi \cite{Ryu:2006bv} provided a precise dictionary between entropy and geometrical quantities in the context of AdS/CFT. The areas of codimension $2$ extremal surfaces in AdS$_{n+1}$ are associated with fine-grained entropies in the CFT.\footnote{What we refer to as fine-grained entropy is the von Neumann entropy, which is bounded from above by the usual thermodynamic entropy. We will define this notion in Appendix \ref{app:entanglement_entropy}.} Similarly to the thermodynamic entropy of black holes, this relation must be corrected in semiclassical regimes, by replacing the geometrical entropy of a codimension $2$ surface $B$ with the fine-grained generalized entropy \cite{Faulkner:2013ana}:
\begin{equation}
\label{eq:Sgen}
    S_{\rm gen}(B) = \frac{A(B)}{4\hat{G}} + S(B),
\end{equation}
where $A(B)$ is the area of $B$ and $S(B)$ is the fine-grained entropy of the bulk fields across $B$.

The notion of generalized entropy is particularly useful as it is divergence-free. Both the Newton's constant $\hat{G}$ and the fine-grained entropy are cutoff-dependent. On the other hand, several pieces of evidence \cite{Susskind:1994sm, Jacobson:1994iw, Frolov:1996aj} show that the generalized entropy is a cutoff-independent (divergence-free) quantity. This highlights the idea that the generalized entropy contains information about the complete theory of quantum gravity. From another perspective, it has been shown in \cite{Chandrasekaran:2022cip} that gravity and quantum field theory must be coupled in order to define an algebra of observables in which entropy is properly defined. In particular, semiclassical entropy in this theory is given by the generalized entropy \eqref{eq:Sgen}.

Motivated by these observations, Strominger and Thompson \cite{Strominger:2003br} conjectured that applying the modification $A/4\hat{G} \rightarrow S_{\rm gen}$ to the classical Bousso bound leads to a quantum version of the Bousso bound that holds at the semiclassical level. They proved the validity of their conjecture in two vacuum states of the Russo–Susskind–Thorlacius (RST) model \cite{Russo:1992ax}, a two-dimensional model of evaporating black hole in AdS. An alternative quantum Bousso bound was proposed in \cite{Bousso:2015mna} by Bousso, Fisher, Leichenauer and Wall, which also uses the notion of generalized entropy but has not been proven yet.

In this paper, we consider these two quantum Bousso bounds in the conformal vacua of semiclassical Jackiw-Teitelboim (JT) gravity \cite{Jackiw:1984je,Teitelboim:1983ux}, both in de Sitter and Anti-de Sitter backgrounds. Our motivation is twofold. First, Strominger and Thompson considered a background with negative cosmological constant. Since the Bousso bound has been central in recent developments of holography in cosmology and de Sitter space \cite{Fischler:1998st, Bousso:1999cb,Hartman:2020khs ,Susskind:2021omt, Franken:2023pni, Franken:2023jas}, we would like to study its quantum versions on a background with a positive cosmological constant as well. Second, the proof of \cite{Strominger:2003br} is only valid in two specific conformal vacua of the RST model, and the quantum Bousso bound conjectured in \cite{Bousso:2015mna} was not shown yet. Hence, we would like to provide a proof of both bounds that is valid in any conformal vacuum states of JT gravity.
\paragraph{Outline}~~\\
~\\~
After a short introduction to JT gravity, the goal of \Sect{sect:dS_JT} is to present a new result concerning entanglement entropy in de Sitter JT gravity, that will be important in the rest of the paper. Since this result is specific to the de Sitter background, and to provide some context, we introduce JT gravity in de Sitter space, which has two different models. The first one is a $\mathbb{Z}_2$-orbifold of dS$_2$, and the second one is a two-dimensional reduction of Schwarzschild-de Sitter space. These models have been investigated in the literature \cite{Cotler:2019nbi, Aalsma:2019rpt, Maldacena:2019cbz, Sybesma:2020fxg, Balasubramanian:2020xqf, Aalsma:2021bit, Pedraza:2021cvx, Kames-King:2021etp, Svesko:2022txo, Piao:2023vgm} to study islands and information recovery in dS. We first review them in classical gravity, and then include in Section \ref{sect:semi_classical_JT} semiclassical corrections by coupling gravity to a two-dimensional CFT with central charge $c$. This semiclassical analysis is independent of the background, and applies to both AdS and dS versions of JT gravity. In \Sect{sec:entropy_half}, we discuss the half reduction model with reflective boundaries and provide the formula for the von Neumann entropy of a Cauchy slice ending on one boundary. This result is important as it tackles an issue that has been overlooked in the literature so far: in the half reduction model, left and right-moving modes are correlated, modifying the usual entropy formula. In particular, the fine-grained entropy of a Cauchy slice bounded by a point and the boundary should not depend on the endpoint at the boundary. Although we confine ourselves to the Bousso bound in this paper, we expect that this formula may lead to developments or modifications of existing results in the half reduction model of de Sitter JT gravity. We will highlight that the properties of semiclassical JT gravity that are used in this paper, apart from the discussion of Section \ref{sec:entropy_half}, do not depend on the background. We do not describe further the geometry of Anti-de Sitter JT gravity, see \eg \cite{Mertens:2022irh} for a review.  

The classical Bousso bound in the framework of JT gravity is introduced in \Sect{sec:lightsheets}. We start by presenting the notions of light-sheet, expansion, and energy conditions in JT gravity. Following the arguments of \cite{Strominger:2003br}, we then prove the Bousso bound in classical JT gravity. 

Moving to the semiclassical regime where the Null Energy Condition (NEC) may be violated, we introduce the notion of quantum light-sheet, together with the Quantum Null Energy Condition (QNEC) and the Quantum Bousso Bound (QBB) of Strominger and Thompson in Section \ref{sec:ST}. We then prove the Strominger-Thompson QBB in JT gravity, in the infinite family of vacuum states of the CFT. This is carried out by assuming the same entropy conditions as the ones considered in \cite{Strominger:2003br}, as introduced in Section \ref{sec:gen_entropy}. Investigating quantitatively the precise regimes in which those conditions are satisfied would be an interesting question, which however goes beyond the scope of this work. We achieve this proof in two steps. First, we prove a sufficient condition for the bound: the normal ordered stress tensor $\braket{:T_{\mu\nu}:}$ satisfies the inequality
\begin{equation}
    2\pi k^{\mu}k^{\nu}\braket{:T_{\mu\nu}:}\geq S'',
\end{equation}
where $k^{\mu}$ is the tangent vector to the light-sheet and $S''$ is the second derivative of the fine-grained entropy of a Cauchy slice bounded by a point of the light-sheet. The second part of the proof consists of showing that
\begin{equation}
     \mathcal{Q} \equiv 2\pi k^{\mu}k^{\nu}\braket{:T_{\mu\nu}:} - S'' - \frac{6}{c}(S')^2 = 0
\end{equation}
in any conformal vacuum, by verifying that $\mathcal{Q}$ is invariant under a change of vacuum state and that it vanishes in the Bunch-Davies/Hartle-Hawking state. In other words, the sufficient condition is always satisfied. This last property is a stronger version of the QNEC showed by Wall \cite{Wall:2011kb} in arbitrary two-dimensional conformal vacua, $2\pi k^{\mu}k^{\nu}\braket{T_{\mu\nu}} - S'' - \frac{6}{c}(S')^2 \geq 0$. In addition to investigating the Strominger-Thompson conjecture in a model with either positive or negative cosmological constant, our proof goes beyond that of \cite{Strominger:2003br} in that it applies to all conformal vacua of the matter fields, instead of specific examples of conformal vacuum states. We then conclude by comparing the Strominger-Thompson bound to two other proposals of QBB \cite{Bousso:2014sda, Bousso:2015mna}, one of which emerges from a semiclassical version of the focussing theorem: the Quantum Focussing Conjecture (QFC) \cite{Bousso:2015mna}. In particular, the Bousso-Fisher-Leichenauer-Wall bound \cite{Bousso:2015mna} follows from an analogous derivation, without requiring the QFC. Finally, possible future works are briefly discussed in Section \ref{sec:_conclusion}.

In Appendix \ref{app:entanglement_entropy}, we review fine-grained entropy in two-dimensional CFTs, and we derive an entropy formula in the case where the system is spatially bounded by reflecting boundaries. We also derive the transformation laws of the entropy under a change of vacuum state. Lastly, the formalism of light-sheet, expansion, and energy conditions in arbitrary dimension is reviewed in Appendix \ref{sec:hydro_regime}.

\section{Jackiw-Teitelboim gravity}
\label{sect:dS_JT}

In this section, we introduce two-dimensional JT gravity and present in \Sect{sec:entropy_half} a semiclassical entropy formula in the half reduction model of de Sitter JT gravity. To motivate this result, we review in Section \ref{sec:JTdS_cl} relevant aspects of the classical geometry of de Sitter JT gravity, in the spirit of \cite{Cotler:2019nbi, Maldacena:2019cbz, Sybesma:2020fxg, Aalsma:2021bit, Pedraza:2021cvx, Kames-King:2021etp,Svesko:2022txo}. We then introduce semiclassical JT gravity in \Sect{sect:semi_classical_JT}, a discussion valid both on an AdS or dS background. Apart from the entropy formula presented in \Sect{sec:entropy_half} that is specific to de Sitter space, the properties of JT gravity that will be used in \Sects{sec:lightsheets} and \ref{sec:ST} are independent of the background. 

The JT gravity action comes from a spherical reduction, from $n+1$ to $2$ dimensions, of the Einstein-Hilbert action with cosmological constant $\hat{\Lambda}$ \cite{Jackiw:1984je, Teitelboim:1983ux}:
\begin{equation}
\label{eq:JTaction}
    I_{\rm EH}=\frac{1}{16\pi \hat{G}}\int_{\hat{\mathcal{M}}}\d^{n+1}X\sqrt{-\hat{g}}\left[\hat R-2\hat{\Lambda}\right] + \frac{1}{8\pi \hat{G}}\int_{\partial\hat{\mathcal{M}}} \d^{n+1}Y\sqrt{-\hat{h}}\hat{K}.
\end{equation}
We denote by $\{X^{M},~M=0,...,n\}$ the coordinates on the $(n+1)$-dimensional manifold $\hat{\mathcal{M}}$, $\hat{g}_{MN}$ the metric tensor on $\hat{\mathcal{M}}$ and $\hat{g}$ its determinant, $\hat{R}$ the Ricci scalar and $\hat G$ the $(n+1)$-dimensional Newton's constant. The second term in \eqref{eq:JTaction} is the Gibbons-Hawking-York boundary term, defined on the $n$-dimensional boundary $\partial\hat{\mathcal{M}}$ of $\hat{\mathcal{M}}$. $\{Y^{M},~M=0,...,n-1\}$ are the coordinates on $\partial\hat{\mathcal{M}}$, $\hat h_{MN}$ is the induced metric on $\partial\hat{\mathcal{M}}$, with $\hat{h}$ its determinant and $\hat{K}$ the trace of its extrinsic curvature.

The $(n+1)$-dimensional (Anti-)de Sitter space, (A)dS$_{n+1}$, is the maximally symmetric solutions with (negative) positive cosmological constant of the equations of motion derived from \eqref{eq:JTaction}. The radius of curvature $l_n$ of (A)dS$_{n+1}$ is related to $\hat{\Lambda}$ by $\hat{\Lambda}=\pm n(n-1)/(2l_n^2)$, with the positive and negative sign corresponding to dS$_{n+1}$ and AdS$_{n+1}$, respectively. We then consider the spherical reduction of the action \eqref{eq:JTaction} with the metric ansatz \cite{Svesko:2022txo}
\begin{equation}
\label{eq:sphred}
    \d s^2 = \hat{g}_{MN}\d X^M\d X^N = g_{\mu\nu}(x) \d x^{\mu}\d x^{\nu} + l_n^2\Phi^{2/(n-1)}(x)\d\Omega_{n-1}^2.
\end{equation}
We will call $\mathcal{M}$ the two-dimensional manifold described by the metric $g_{\mu\nu}$ with coordinates $\{x^{\mu},~\mu=0,1\}$, such that $\hat{\mathcal{M}} =\mathcal{M}\cross\mathbb{S}^{n-1}$. $\Phi$ is called the dilaton, which encodes the size of the $(n-1)$-dimensional compact space $\mathbb{S}^{n-1}$. The dimensional reduction of \eqref{eq:JTaction} then gives the two-dimensional action
\begin{align}\label{eq:dim_red_action}
I_{\rm EH}&=\frac{1}{16\pi G}\int_{\mathcal{M}}\d^2 x\sqrt{-g}\left[\Phi R-2\hat{\Lambda} \Phi+\frac{(n-2)(n-1)}{l_n^2}\Phi^{\frac{n-3}{n-1}}+\frac{n-2}{n-1}\frac{(\nabla \Phi)^2}{\Phi}\right]\nonumber\\
&+\frac{1}{8\pi G}\int_{\partial\mathcal{M}}\d y \sqrt{-h}\Phi K,
\end{align}
where $R$ is the two-dimensional Ricci scalar, $\nabla$ the covariant derivative compatible with the metric $g_{\mu\nu}$, and $K$ the trace of the extrinsic curvature of the boundary $\partial\mathcal{M}$ of $\mathcal{M}$. The two-dimensional Newton's constant $G$ is given by 
\begin{equation}\label{eq:2d_Newton_cste}
\frac{1}{G} = \frac{S_{n-1}(l_n)}{\hat{G}},
\end{equation}
with $S_{n-1}(l_n)=2\pi^{n/2}~l_n^{n-1}/\Gamma(n/2)$ the surface area of the $(n-1)$-sphere of radius $l_n$. 
As we will discuss in the next section, this dimensional reduction for $\hat\Lambda>0$ with $n=2$ and $n\geq 3$ leads to two distinct versions of de Sitter JT gravity, respectively called half and full reduction models. While having the same equations of motion, these two models have very different effective geometries.

\subsection{Two-dimensional (Anti-)de Sitter space}
\label{sec:JTdS_cl}

Let us consider $\hat{\Lambda}>0$. The dimensionally reduced action \eqref{eq:dim_red_action} simplifies considerably for $n=2$. In this case, the scalar potential and the kinetic term of the dilaton vanish, yielding the action
\begin{equation}
\label{eq:actionh}
    I^{\rm half}_{\rm JT}=\frac{1}{16\pi G}\int_{\mathcal{M}} \d^2x\sqrt{-g}~\Phi(R-2\Lambda)+ \frac{1}{8\pi G}\int_{\partial\mathcal{M}} \d y \sqrt{-h}\Phi K,
\end{equation}
with $\Lambda=1/l_2^2$. This is the action of the \emph{half reduction model} of de Sitter JT gravity. The situation for $n\geq 3$ is more involved. The kinetic term for the dilaton can be removed by a suitable Weyl rescaling of the metric $g_{\mu\nu}\rightarrow \frac{1}{n}\Phi^{-\frac{n-2}{n-1}}g_{\mu\nu}$, but a potential will still remain \cite{Svesko:2022txo}. Expanding around the minimum of the potential $\phi_0=1$ gives the action of the \emph{full reduction model} of de Sitter JT gravity:
\begin{align}
\label{eq:actionf}
    I^{\rm full}_{\rm JT}=\frac{1}{16\pi G}\int_{\mathcal{M}} \d^2x\sqrt{-g}~((\phi_0+\phi)R-2\Lambda\phi)+ \frac{1}{8\pi G}\int_{\partial\mathcal{M}} \d y \sqrt{-h}(\phi_0+\phi) K,
\end{align}
with $\Lambda=1/l_n^2$. The dilaton $\Phi(x) = \phi_0 + \phi(x)$ can be seen as an expansion around the Nariai geometry $\Phi=\phi_0$, with $\phi$ describing the deviation (which must be small compared to $\phi_0$) away from the Nariai solution, in analogy with AdS JT gravity \cite{Svesko:2022txo}\footnote{The Nariai geometry describes the largest black hole fitting inside de Sitter space \cite{Nariai}. For the full reduction model to explicitly derive from the Nariai geometry, one needs to modify the ansatz \eqref{eq:sphred} into
\begin{equation}
\label{eq:Nariai}
    \d s^2 = \hat{g}_{MN}\d X^M\d X^N = g_{\mu\nu}(x) \d x^{\mu}\d x^{\nu} + r_N^2\Phi^{2/(n-1)}(x)\d\Omega^{n-1},
\end{equation}
where $r_N=\sqrt{\frac{n-2}{n}}l_n$ is the Nariai radius.}. From the two-dimensional point of view, $\phi_0$ is a topological term which does not modify the equations of motion, so that the explicit form of $\phi$ is the same in both models. However, it allows $\phi$ to take negative values, changing the topology of the two-dimensional space. 

The holographic principle assigns to any codimension $2$ surface an entropy given by its area divided by $4\hat{G}$. A point in $\mathcal{M}$ corresponds, from the $(n+1)$-dimensional point of view, to an $(n-1)$-sphere at constant $x$, with area $S_{n-1}(l_n)\Phi(x)$ as follows from the metric ansatz \eqref{eq:sphred}. Using \eqref{eq:2d_Newton_cste}, we get
\begin{equation}
\label{area}
    \frac{S_{n-1}(l_n)\Phi(x)}{4\hat{G}} = \frac{\Phi}{4G},
\end{equation}
which brings us to interpret $\Phi$ as the ``area'' of a point in JT gravity, and therefore to impose $\Phi\geq 0$. In the half reduction, this leads to $\phi\geq 0$, while in the full reduction, negative values of $\phi$ are allowed as long as $\phi\geq-\phi_0$.

The two-dimensional bulk dynamics is usually studied in Kruskal coordinates $(x^+,x^-)$ in the conformal gauge, where the metric reads
\begin{equation}\label{eq:metric_conf_gauge}
\d s^2=-e^{2\omega(x^+,x^-)}\d x^+ \d x^-.
\end{equation}
The non-vanishing Christoffel symbols associated to this metric are $\Gamma_{++}^+=2\partial_+\omega$, $\Gamma_{--}^-=2\partial_{-}\omega$, while the Ricci scalar is given by $R=8e^{-2\omega}\partial_+\partial_-\omega$. Varying the action with respect to the dilaton yields the equation of motion for the metric:
\begin{equation}
R=2\Lambda\quad \Leftrightarrow \quad \partial_+\partial_-\omega=\frac{\Lambda}{4}e^{2\omega},
\end{equation}
hence fixing the background geometry to be de Sitter. This is solved by
\begin{equation}\label{eq:class_sol_omega}
e^{2\omega(x^+,x^-)}=\frac{4}{(1-\Lambda x^+x^-)^2}.
\end{equation}
On the other hand, varying the action with respect to the inverse metric yields the equation of motion for $\phi$,
\begin{equation}\label{eq:classical_eom}
(g_{\mu\nu}\nabla^2-\nabla_{\mu}\nabla_{\nu}+g_{\mu\nu}\Lambda)\phi=0.
\end{equation}
In Kruskal coordinates and in the conformal gauge, the latter equation splits into its off-diagonal and diagonal parts:
\begin{eqnarray}
2\partial_+\partial_-\phi-\Lambda e^{2\omega}\phi&=& 0,\\
-\partial_{\pm}^2\phi+2\partial_{\pm}\omega\partial_{\pm}\phi&=&0,
\end{eqnarray}
Using \eqref{eq:class_sol_omega}, these equations are solved by
\begin{equation}
\label{eq:vacdil}
\phi_*(x^+,x^-)=\phi_r\frac{1+\Lambda x^+x^-}{1-\Lambda x^+x^-},
\end{equation}
with $\phi_r>0$ a constant. In the half reduction model, $\phi_r$ can be normalized to one by matching the Wald entropy associated to the horizon with the Gibbons-Hawking entropy of dS$_3$ \cite{Svesko:2022txo}. We will keep it general in this work.

While pure two-dimensional gravity is non-dynamical, the geometry of spacetime in JT gravity is described by the dilaton field. In particular, the condition $\Phi\geq 0$ implies
\begin{equation}
    -\frac{1}{\Lambda} \leq x^+x^- \leq \frac{1}{\Lambda},
\end{equation}
in the half reduction case, where $x^+x^-=1/\Lambda$ correspond to the past and future null infinity $\cal{J}^{\pm}$ and $x^+x^-=-1/\Lambda$ correspond to the spatial boundaries.
\begin{figure}[t!]
\centering
\begin{tikzpicture}
\begin{scope}[transparency group]
\begin{scope}[blend mode=multiply]
\path
       +(3,3)  coordinate (IItopright)
       +(-3,3) coordinate (IItopleft)
       +(3,-3) coordinate (IIbotright)
       +(-3,-3) coordinate(IIbotleft)
      
       ;
\draw (IItopleft) --
          node[midway, above, sloped] {{\footnotesize$\cal{J}^+$}}
          node[midway, below, sloped] {{\footnotesize$\Phi=+\infty,\ x^+x^-=1/\Lambda$}}
      (IItopright) --
          node[midway, above,sloped] {Antipode}
          node[midway, below,sloped] {{\footnotesize $\Phi=0,\ x^+x^-=-1/\Lambda$}}
      (IIbotright) -- 
          node[midway, above, sloped] {{\footnotesize $\Phi=+\infty,\ x^+x^-=1/\Lambda$}}
          node[midway, below, sloped] {{\footnotesize $\cal{J}^-$}}
      (IIbotleft) --
          node[midway, above , sloped]  {Pode}
          node[midway, below , sloped] { {\footnotesize $\Phi=0,\ x^+x^-=-1/\Lambda$}}
      (IItopleft) --  cycle;

\draw (IItopleft) -- node[midway, above, sloped] {{\footnotesize $\Phi=\phi_r$}}(0,0) -- (IItopright);
\draw (IIbotleft) -- (0,0) -- (IIbotright);

\draw (IItopleft) -- (-3,3) ;
\draw (IItopright) -- (3,3) ;
\draw (IItopright) -- (3,3) ;

\fill[fill=blue!20] (-3,3) -- (0,0) -- (-3,-3);
\fill[fill=blue!20] (3,3) -- (0,0) -- (3,-3);

\end{scope}
\end{scope}
\end{tikzpicture} 
    \caption{\footnotesize Penrose diagram for two-dimensional de Sitter space in the half reduction model. Any spacelike slice is a segment whose boundaries are called pode and antipode; so that the full dS$_2$ spacetime in the half reduction has two timelike boundaries depicted by the two vertical lines. The diagonal lines are the past and future cosmological horizons for an observer at the pode and antipode, which delimit the two static patches depicted in blue. Inside each static patch, the dilaton varies from $\Phi=0$ on the pode/antipode to the constant positive value $\Phi=\phi_r$ on the cosmological horizons. It diverges to $\Phi\rightarrow +\infty$ on the past and future infinity $\cal{J}^{\pm}$. \label{fig:Penrose_diag_half}}
\end{figure}
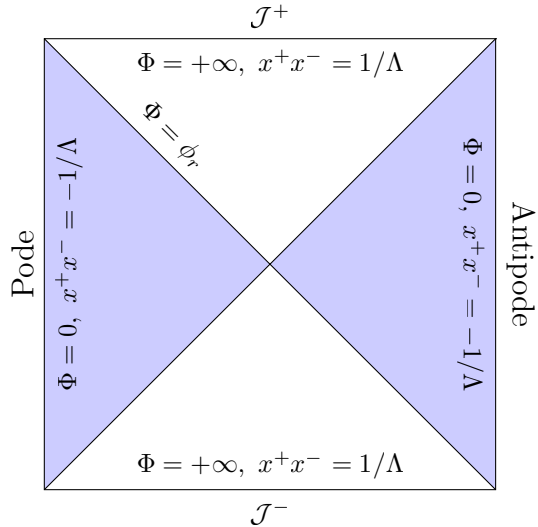
From the higher dimensional perspective, these timelike boundaries correspond to the poles of dS$_3$, and we still call them poles (pode and antipode) in the two-dimensional setup. Spatial slices are segments bounded by the poles, where comoving observers are located. Each observer has an associated cosmological horizon located at $x^+ x^-=0$, which bounds their respective causal patches. The Penrose diagram for two-dimensional de Sitter space in the half reduction model is depicted in \Fig{fig:Penrose_diag_half}, which can be seen as a $\mathbb{Z}_2$-orbifold of dS$_2$.\footnote{This comes from the condition $\Phi\geq 0$ \cite{Aalsma:2021bit}.}.

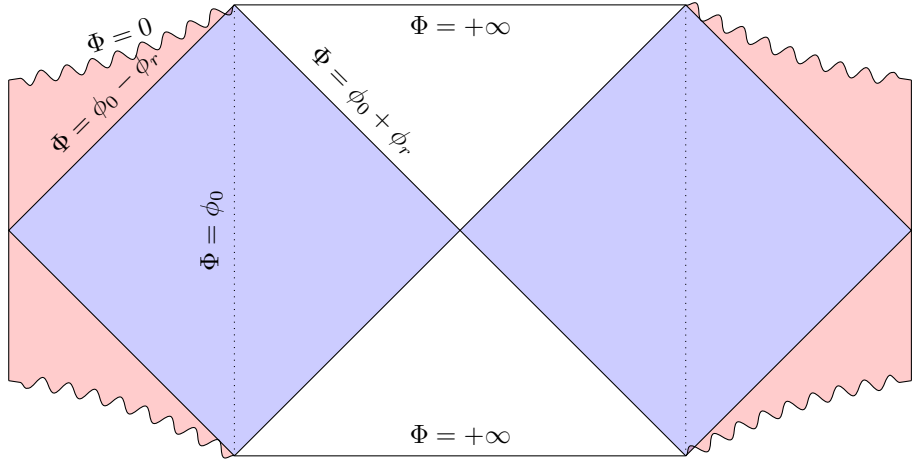
\begin{figure}[h!]
\centering
\begin{tikzpicture}
\begin{scope}[transparency group]
\begin{scope}[blend mode=multiply]
\path
       +(6,3)  coordinate (IItopright)
       +(0,3)  coordinate (IItopcenter)
       +(-6,3) coordinate (IItopleft)
       +(0,3)  coordinate (IIbotcenter)
       +(6,-3) coordinate (IIbotright)
       +(-6,-3) coordinate(IIbotleft)
      
       ;

\draw (-3,3) -- node[midway, below, sloped] {{\footnotesize$\Phi=+\infty$}}(3,3) ;
\draw (-3,-3) -- node[midway, above, sloped] {{\footnotesize$\Phi=+\infty$}}(3,-3) ;
\draw (-6,-2) -- (-6,2) ;
\draw (6,-2) -- (6,2) ;

\draw (-3,3) -- node[midway, above, sloped] {{\footnotesize $\Phi=\phi_0-\phi_r$}}(-6,0) -- (-3,-3) ;
\draw (3,3) -- (6,0) -- (3,-3) ;
\draw (-3,3) -- (3,-3) ;
\draw (-3,-3) -- (3,3) ;

\fill[fill=blue!20] (-6,0) -- (-3,3) -- node[midway, above, sloped] {{\footnotesize $\Phi=\phi_0+\phi_r$}} (0,0) -- (-3,-3);
\fill[fill=blue!20] (6,0) -- (3,3) -- (0,0) -- (3,-3);

\fill[fill=red!20] (-3,3) decorate[decoration=snake] {to[bend left=10] (-6,2)} -- (-6,0);
\fill[fill=red!20] (3,3) decorate[decoration=snake] {to[bend left=-10] (6,2)} -- (6,0);
\fill[fill=red!20] (-3,-3) decorate[decoration=snake] {to[bend left=-10] (-6,-2)} -- (-6,0);
\fill[fill=red!20] (3,-3) decorate[decoration=snake] {to[bend left=10] (6,-2)} -- (6,0);

\draw[dotted] (-3,-3)--node[midway, above, sloped] {{\footnotesize $\Phi=\phi_0$}}(-3,3);
\draw[dotted] (3,3)--(3,-3);

\draw[decorate,decoration=snake] (-3,3) to[bend left=10] node[midway, above, sloped] {{\footnotesize $\Phi=0$}}(-6,2);
\draw[decorate,decoration=snake] (-3,-3) to[bend left=-10] (-6,-2);
\draw[decorate,decoration=snake] (3,3) to[bend left=-10] (6,2);
\draw[decorate,decoration=snake] (3,-3) to[bend left=10] (6,-2);

\end{scope}
\end{scope}
\end{tikzpicture}
    \caption{\footnotesize Penrose diagram for two-dimensional de Sitter space in the full reduction model. The left and right edges are identified, so that any complete spacelike slice is topologically a circle. The two vertical dashed lines depict the pode and the antipode where the dilaton takes the constant value $\Phi=\phi_0$, and the two blue shaded regions their associated static patches. The red shaded regions are the ``black hole'' interiors, with past and future singularities depicted by the wavy lines, where $\Phi=0$. The white regions are the exterior regions of de Sitter. The white and blue regions are covered by Kruskal coordinates $x^{\pm}$. The geometry contains two horizons associated with an observer sitting at the pode/antipode: a cosmological horizon between the static patch and the de Sitter exterior region, where the dilaton takes the constant value $\Phi=\phi_0+\phi_r$, and a black hole horizon between the static patch and the black hole interior, where the dilaton takes the constant value $\Phi=\phi_0-\phi_r$.  }
    \label{fig:Penrose_diag_full}
\end{figure}

In the full reduction case, $x^+ x^-$ is still bounded from above by $1/\Lambda$, corresponding to $\cal{J}^{\pm}$. However, it is no longer bounded from below, with $x^+x^- \rightarrow -\infty$ corresponding to the black hole horizons. In fact, the full reduction model can be maximally extended such that it is periodic in space and includes the black hole region of the Schwarzschild-de Sitter space \cite{Svesko:2022txo}, see \Fig{fig:Penrose_diag_full}. The Kruskal coordinates $(x^+,x^-)$ cover the entire dS$_2$ spacetime in the half reduction model, and the blue and white region of \Fig{fig:Penrose_diag_full} in the full reduction model. As we will see in the next section, coupling matter to JT gravity induces a backreaction on the dilaton, hence changing the effective geometry.

We will highlight in Section \ref{sec:entropy_half} the importance of distinguishing between the half and full reduction models in de Sitter space when considering entanglement entropies. On the other hand, the interest towards Anti de Sitter JT gravity is older and has been extensively studied in the literature. In particular, Anti-de Sitter JT gravity follows from the dimensional reduction of the action \ref{eq:actionf} describing near-extremal AdS black hole, similarly to the full reduction model of de Sitter JT gravity obtained from the dimensional reduction of Schwarzschild-dS black hole. 
Hence, as one can see from equations \ref{eq:sphred} and \ref{eq:dim_red_action}, the AdS and dS JT gravity actions only differ by the sign of the cosmological constant $\Lambda$, and so does the dilaton equation of motion \eqref{eq:classical_eom}. Crucially, we will see in \Sects{sec:lightsheets} and \ref{sec:ST} that the $\Lambda$-dependent term of \eqref{eq:classical_eom} does not enter in the proofs of the classical and quantum Bousso bounds, so that the results presented in these sections are valid both in AdS and dS JT gravity.  
The semiclassical analysis of Section \ref{sect:semi_classical_JT} is also independent of the dS or AdS background, while the discussion of Section \ref{sec:entropy_half} concerns only de Sitter JT gravity. Therefore, we do not go into further detail concerning the geometry of Anti-de Sitter JT gravity, and refer the reader to \eg \cite{Mertens:2022irh} for a review.

\subsection{Semiclassical JT gravity}
\label{sect:semi_classical_JT}

Semiclassical JT gravity is obtained by coupling the geometrical action $I_{\rm JT}$ introduced above to a $2d$ CFT with central charge $c$, described by an action $I_{\rm CFT}$. In this paper, we work in the semiclassical limit where $c\rightarrow \infty$, $G\rightarrow 0$, while keeping $cG$ fixed. In the large $c$ limit, the backreaction is fully captured by the (non-local) $1$-loop Polyakov action \cite{Polyakov:1981rd}
\begin{equation}
    I_{\rm Polyakov} = - \frac{1}{16\pi G}\int \d^2x\sqrt{-g}\left[\frac{c}{48}R\frac{1}{\nabla^2}R\right].
\end{equation}
The total action is thus given by 
\begin{equation}\label{eq:2d_action}
I = I_{\rm JT} + I_{\rm Polyakov} + I_{\rm CFT},
\end{equation}
where $I_{\rm JT}=I_{\rm JT}^{\rm half}$ or $I_{\rm JT}=I_{\rm JT}^{\rm full}$, depending on the JT model under consideration. In this work, we will consider a CFT composed of $c$ non-interacting scalar fields $\psi_i$,
\begin{equation}\label{eq:CFT_action}
    I_{\rm CFT} = \frac{1}{16\pi G} \int \d x^2 \sqrt{-g}\left[-\frac{1}{2} \sum_{i=1}^c\left( \nabla \psi_i \right)^2  \right].
\end{equation}
The nature of the CFT action will not be of importance here, but will be necessary in the introduction of entanglement entropies.

In the half reduction model, the conformal field theory is defined on a curved background \eqref{eq:metric_conf_gauge} with spatial boundaries $x^+x^-=-1/\Lambda$. In particular, to provide a physically meaningful picture, consistent with the higher dimensional one, we should impose reflecting boundary conditions:
\begin{equation}
    \left.\psi_i\right|_{x^+x^-=-1/\Lambda}=0.
\end{equation}
We will see in Section \ref{sec:entropy_half} that the importance of this boundary condition has been overlooked in the past, and that it has crucial consequences when computing entanglement entropies. 

At the semiclassical level, the equation of motion for $\phi$ reads
\begin{equation}
\label{eq:dilaton_eom}
(g_{\mu\nu}\nabla^2-\nabla_{\mu}\nabla_{\nu}+g_{\mu\nu}\Lambda)\phi=8\pi GT_{\mu\nu},
\end{equation}
or equivalently in conformal gauge:
\begin{eqnarray}
2\partial_+\partial_-\phi-\Lambda e^{2\omega}\phi&=&16\pi G T_{+-},\\
\label{eq:T++_eom}
-\partial_{\pm}^2\phi+2\partial_{\pm}\omega\partial_{\pm}\phi&=&8\pi G T_{\pm \pm}.\label{eom++}
\end{eqnarray}
The stress-energy tensor of the matter CFT can be written as 
\begin{equation}
T_{\mu\nu}=\tau_{\mu\nu}+T_{\mu\nu}^{\rm qu},
\end{equation}
where $\tau_{\mu\nu}$ is the (state-independent) flat space contribution to the stress tensor, coming from the variation of $I_{\rm CFT}$\footnote{\label{fn:stresstensor}With the CFT action \eqref{eq:CFT_action}, we have $
\tau_{\mu\nu}=1/(16\pi G)\sum_{i=1}^c\left(\nabla_{\mu}\psi_i\nabla_{\nu}\psi_i-\frac{1}{2}g_{\mu\nu}(\nabla\psi_i)^2\right)$, or in lightcone coordinates, $\tau_{\pm\pm}=(1/16\pi G)\sum_{i=1}^c(\partial_{\pm}\psi_i)^2, \tau_{+-}=0$, but we will not need this explicit expression in this paper.}, and $T_{\mu\nu}^{\rm qu}$ its quantum mechanical part coming from the variation of the Polyakov action $I_{\rm Polyakov}$. This term captures the (state-independent) conformal anomaly that the stress-energy tensor acquires on a curved spacetime background, which reads:
\begin{equation}\label{eq:conf_anomaly}
g^{\mu\nu}\braket{T_{\mu\nu}}=\frac{c}{24\pi}R ~\Longrightarrow~ \braket{T_{+-}}=-\frac{c}{12\pi}\partial_+\partial_-\omega.
\end{equation}
In the same spirit as in \cite{Christensen:1977jc}, one can use \eqref{eq:conf_anomaly} to solve the conservation equation
\begin{equation}\label{eq:conservation_eq}
    \nabla_{\mu}\braket{T^{\mu\nu}}=0,
\end{equation}
and get the components of the stress-energy tensor:
\begin{eqnarray}
\braket{T_{\pm\pm}(x^{\pm})}&=&\braket{\tau_{\pm\pm}}+\frac{c}{12\pi}\left(\partial_{\pm}^2\omega-\partial_{\pm}\omega\partial_{\pm}\omega\right) -\frac{c}{24\pi}t_{\pm}(x^{\pm}),\\
\braket{T_{+-}(x^+,x^-)}&=&-\frac{c}{12\pi}\partial_+\partial_-\omega.
\end{eqnarray}
The functions $t_{\pm}(x^{\pm})$ are functions of integration that arise in the integration of the continuity equation \eqref{eq:conservation_eq}. As we will see in the following, they are state-dependent and characterize the choice of vacuum. Using the classical solution \eqref{eq:class_sol_omega}, these expressions simplify to
\begin{eqnarray}
\braket{T_{\pm\pm}(x^{\pm})}&=& \braket{\tau_{\pm\pm}}-\frac{c}{24\pi}t_{\pm}(x^{\pm}),\\
\braket{T_{+-}(x^+,x^-)}&=&-\frac{c}{12\pi}\frac{\Lambda}{(1-\Lambda x^+x^-)^2}.
\end{eqnarray}
In terms of the normal ordered stress tensor $:T_{\pm\pm}(x^{\pm}):~\equiv~ T_{\pm\pm}(x^{\pm}) - \tau_{\pm\pm}(x^{\pm})$, we thus have
\begin{equation}
    \braket{:T_{\pm\pm}(x^{\pm}):}=-\frac{c}{24\pi}t_{\pm}(x^{\pm}).
\end{equation}
These expressions are only valid in Kruskal coordinates, as the stress-energy tensor transforms non-trivially under conformal transformation $x^{\pm} \rightarrow y^{\pm}(x^{\pm})$. In particular, $\tau_{\pm\pm}$ transforms as a rank-$2$ tensor, 
\begin{equation}\label{eq:transtau}
    \tau_{\pm\pm}(y^{\pm}) = \left(\frac{\d x^{\pm}}{\d y^{\pm}}\right)^2 \tau_{\pm\pm}(x^{\pm}),
\end{equation}
while the functions $t_{\pm}$ obey the anomalous transformation law:
\begin{equation}
\label{eq:transft}
t_{\pm}(y^{\pm})=\left(\frac{\d x^{\pm}}{\d y^{\pm}}\right)^2 t_{\pm}(x^{\pm})+\{x^{\pm},y^{\pm}\},
\end{equation}
where $\{x^{\pm},y^{\pm}\}$ is the Schwarzian derivative defined by
\begin{equation}
\{x^{\pm},y^{\pm}\}=\frac{\dddot x^{\pm}}{\dot x^{\pm}}-\frac{3}{2}\left(\frac{\ddot x^{\pm}}{\dot x^{\pm}}\right)^2,
\end{equation}
with $\dot x^{\pm}=\d x^{\pm}/\d y^{\pm}$. This yields the following transformation law for the stress tensor:
\begin{equation}
\label{eq:transfT}
\braket{T_{\pm\pm}(y^{\pm})}=\left(\frac{\d x^{\pm}}{\d y^{\pm}}\right)^2 \braket{T_{\pm\pm}(x^{\pm})}-\frac{c}{24\pi}\{x^{\pm},y^{\pm}\}.
\end{equation}

A \emph{vacuum state}, defined as a state where $\braket{:T_{\pm\pm}(y^{\pm}):}=0$ in some coordinate system $y^{\pm}$, therefore corresponds to a state in which the function $t$ vanishes in the coordinates $y^{\pm}$. From this, one sees that $\tau_{\pm\pm}$ corresponds to the vacuum expectation value of $T_{\pm\pm}(y^{\pm})$ in the vacuum state defined with respect to the coordinates $y^{\pm}$, where $t_{\pm}(y^{\pm})=0$. All vacuum states are related by conformal transformations of the lightcone coordinates in which they are defined. We also refer to them as \emph{conformal vacua}.

\subsection{Entropy in the half reduction model}
\label{sec:entropy_half}

Let us consider a two-dimensional CFT with central charge $c$, on a curved background with the metric $\d s^2 =-e^{2\omega_x}\d x^+\d x^-$. In Anti-de Sitter JT gravity or in the full reduction model of de Sitter JT gravity, the von Neumann entropy of a Cauchy slice $\Sigma$, in the vacuum state defined in the coordinates $x^{\pm}$, is given by \cite{Fiola:1994ir}:
\begin{equation}
\label{eq:usualentropy_x}
    S_{x}(\Sigma) = \frac{c}{6}\left(\omega_x(x_1)+\omega_x(x_2)\right) + \frac{c}{12}\ln\left[\frac{(x_2^+-x_1^+)^2(x_2^--x_1^-)^2}{\delta_1^2\delta_2^2}\right],
\end{equation}
where the index $x$ refers to the choice of vacuum state, $(x_1^+,x_1^-)$ and $(x_2^+,x_2^-)$ denote the coordinates of the two endpoints of $\Sigma$, and $\delta_1^2$, $\delta_2^2$ are boost-invariant UV cutoffs at the two endpoints of $\Sigma$, see Appendix \ref{app:entanglement_entropy}. In Anti-de Sitter JT gravity, this formula was shown to match with the Wald entropy in \cite{Pedraza:2021cvx}. 
 
\begin{figure}[h!]
\centering
\begin{tikzpicture}
\begin{scope}[transparency group]
\begin{scope}[blend mode=multiply]
\path
       +(3,3)  coordinate (IItopright)
       +(-3,3) coordinate (IItopleft)
       +(3,-3) coordinate (IIbotright)
       +(-3,-3) coordinate(IIbotleft)
      
       ;

\fill[fill=blue!20] (-1,0) -- (-3,2) -- (-3,-2) --  cycle;
       
\draw (IItopleft) --
      (IItopright) --
      (IIbotright) -- 
      (IIbotleft) --
      (IItopleft) --  cycle;

\draw (IItopleft) -- (0,0) -- (IItopright);
\draw (IIbotleft) -- (0,0) -- (IIbotright);

\draw (IItopleft) -- (-3,3) ;
\draw (IItopright) -- (3,3) ;
\draw (IItopright) -- (3,3) ;

\draw (-1,0) to[bend left=10] node[midway, above, sloped] {{\footnotesize $\Sigma_P$}}(-3,0.5);

\node at (-1,0) [label = above:{$P$}]{};
\node at (-1,0) [circle, fill, inner sep=1.5 pt]{};

\node at (-3,0.5) [label = left:{$A$}]{};
\node at (-3,0.5) [circle, fill, inner sep=1.5 pt]{};

\draw [blue,decorate,decoration=snake](-1,-3) -- (-3,-1) ;
\draw [-{Stealth[length=3mm]},blue,decorate,decoration={snake, post length=2mm}](-3,-1) -- (1,3) ;
\draw [red,decorate,decoration=snake](1,-3) -- (-3,1);
\draw [-{Stealth[length=3mm]},red,decorate,decoration={snake, post length=2mm}](-3,1) -- (-1,3);

\end{scope}
\end{scope}
\end{tikzpicture} 
\caption{\footnotesize
    Penrose diagrams of the half reduction model, with reflective boundary conditions at the spatial boundaries. They imply that the causal ``diamond'' of a Cauchy slice $\Sigma_P$ ending on one boundary is a triangle, depicted by the blue shaded region, and that $S(\Sigma_P)$ is independent of the location of $A$ along the left edge of the triangle. Examples of radiation emanating as left-moving modes from $\cal{J}^-$ are depicted. The red one crosses $\Sigma_P$ as a left-moving mode, while the blue one crosses $\Sigma_P$ as a right-moving mode.
   }
   \label{fig:BD}
\end{figure}
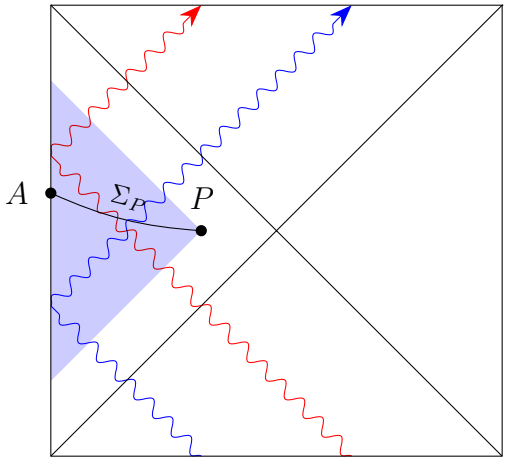

Here, we would like to highlight the fact that this formula is not valid in the half reduction model with reflective boundary conditions. First, we expect the von Neumann entropy of a Cauchy slice covering the full spacetime to vanish, as it is associated with a pure state. Second, the entropy of a slice bounded by a point $P$ and one of the boundaries should not depend on the endpoint at the boundary. Indeed, the reflection of right and left-moving modes on the boundaries ensures that all Cauchy slices joining the point $P$ and the boundary are crossed by the same modes, \ie they have the same entropy. In particular, these slices are related by unitary evolution and share the same causal ``diamond'', as depicted by the blue shaded triangle in Figure \ref{fig:BD}. The entropy formula \eqref{eq:usualentropy_x} does not satisfy either of these two conditions, since the half reduction model is not infinitely extended. 

Considering reflecting boundary conditions at two spatial boundaries parameterized by an arbitrary function $x^-=f_x(x^+)$, the entanglement entropy $S(\Sigma_P)$ associated to a point $P$ located at $(x^+,x^-)$ has been computed in \Appendix{app:reflec_bound} and is given in \eq{eq:ent_bdy}.\footnote{The function $f_x$ is restricted to be a $C^1$ involution on $\mathbb{R}_*$, with $0<\d f_x/\d x^+<\infty$ and $x^+f_x(x^+)\leq 0$, see Appendix \ref{app:reflec_bound}.} In the half reduction model, the function $f_x$ is given by $f_x(x^+)=-1/(\Lambda x^+)$, which gives for the vacuum state defined in the Kruskal coordinates $x^{\pm}$:
\begin{align}\label{eq:ent_bdy_BD}
    S_{x}(\Sigma_P) = \frac{c}{6}\left(\omega_x(x_1)+\omega_x(x_2)\right)+\frac{c}{12}\ln\left[\frac{(1+\Lambda x^+x^-)^2}{\Lambda\delta^2}\right].
\end{align}
This vacuum state is called the Bunch-Davies vacuum. It is analogous to the Minkowski vacuum in flat space or the Hartle-Hawking state of a black hole in thermal equilibrium with its Hawking radiation. Inserting the explicit expression \eqref{eq:class_sol_omega} of $\omega_x$, we find a relation between the entropy and the dilaton solution in the absence of matter \eqref{eq:vacdil}:
\begin{align}\label{eq:entang_entropy}
    S_{x}(\Sigma_P) = \frac{c}{12}\ln\left[\frac{4}{\Lambda\delta^2}\frac{(1+\Lambda x^+x^-)^2}{(1-\Lambda x^+x^-)^2}\right] = \frac{c}{6}\ln\frac{\phi_*}{\phi_r} + \text{constant}.
\end{align}
The reflective boundaries ensure that the state is symmetric with respect to the $x^+$ and $x^-$ coordinates, and that the entropy of $\Sigma_P$ does not depend on the position of its endpoint on the boundary, see Figure \ref{fig:BD}.
This formula can be generalized to any vacuum state, as shown in \eq{eq:transf_bdy} of Appendix \ref{sec:changevac}.

\section{Classical Bousso Bound}
\label{sec:lightsheets}

Let us consider an $(n+1)$-dimensional spacetime $\hat{\mathcal{M}}$ and an arbitrary codimension $2$ spacelike surface $B$ in $\hat{\mathcal{M}}$. A light-sheet of $B$ is a codimension $1$ surface generated by light rays which begin at $B$, extend orthogonally away from $B$ and are of non-positive expansion. Considering the light rays to end at another codimension $2$ spacelike surface $B'$, we denote $L(B-B')$ the part of the light-sheet emanating from $B$ and ending on $B'$. The \emph{Bousso bound}, or \emph{generalized covariant entropy bound}, then states that the coarse-grained entropy $S_{L(B-B')}$ passing through $L(B-B')$ satisfies the inequality \cite{Bousso:1999xy, Bousso:2002ju, Flanagan:1999jp}
\begin{equation}
\label{eq:Bousso}
S_{L(B-B')} \leq \frac{1}{4\hat{G}}(A(B)-A(B')),
\end{equation}
where $\hat G$ is the Newton's constant in $\hat{\mathcal{M}}$, and $A(B)$ and $A(B')$ denote the area of the codimension $2$ surfaces $B$ and $B'$.

The Bousso bound has first been proven in 4 dimensions in \cite{Flanagan:1999jp} under either one of two sets of assumptions. Later, it has been shown that it can actually be derived from simpler assumptions \cite{Bousso:2003kb}. In the following \Sect{sect:BB_dSJT}, the two-dimensional version of the Bousso bound in JT gravity is stated and then proven in \Sect{sect:proof_classical_BB}, following the strategy of \cite{Strominger:2003br}. We refer to \Appendix{sec:hydro_regime} for the definitions of the notions of light-sheet, expansion, and Bousso bound in arbitrary dimension together with their dimensional reduction, although this section can be followed without it. We assume that the classical part $\braket{\tau_{\mu\nu}}$ of the stress-energy tensor satisfies the \emph{Null Energy Condition} (NEC):
\begin{equation}\label{eq:NEC}
    k^{\mu} k^{\nu} \braket{\tau_{\mu\nu}} \geq 0,
\end{equation}
for any null vector $k^{\mu}$. This condition is satisfied by most classical matter fields, and in particular by the classical CFT action \eqref{eq:CFT_action} (see footnote \ref{fn:stresstensor}); though we will see that it can be broken once quantum effects are taken into account. 

\subsection{The Bousso Bound and Energy Conditions}
\label{sect:BB_dSJT}

In this section, we recall the two-dimensional Bousso bound in JT gravity, where the dilaton acts as the area of $0$-dimensional surfaces. For earlier work on the two-dimensional Bousso bound, see \cite{Fukuma:2003uz}.

As discussed at the beginning of \Sect{sect:dS_JT}, de Sitter JT gravity comes from the dimensional reduction of the $(n+1)$-dimensional Einstein-Hilbert action. Considering the metric ansatz \eqref{eq:sphred}, the dimensional reduction of $L(B-B')$ down to two dimensions gives a single light ray bounded by two points $B$ and $B'$, which we will still denote by $L(B-B')$. We call $\lambda$ the affine parameter along $L(B-B')$, normalized such that $\lambda=0$ on $B$ and $\lambda=1$ on $B'$. The future directed vector normal (and tangent) to $L(B-B')$ is defined by
\begin{equation}
    k^{\mu} = \pm \left(\frac{\d}{\d\lambda}\right)^{\mu}.
\end{equation}
The ``$+$'' sign is considered for a future-directed light-sheet, and the ``$-$'' sign for a past-directed light-sheet, so that $k$ is always future-directed. In this paper, the prime will denote the derivative with respect to $\lambda$, \ie $'\equiv \d/\d\lambda=k^{\mu}\nabla_{\mu}$. $\lambda$ being an affine parameter, $k^{\mu}$ must satisfy the geodesic equation \cite{Wald:1984rg}:
\begin{equation}
    k^{\mu}\nabla_{\mu}k^{\nu} = 0.
\end{equation}
The geodesic always propagates either along $x^+$ or $x^-$, such that $k^{\mu}=(k^+,0)$ or $k^{\mu}=(0,k^-)$ respectively. This yields the differential equation
\begin{align}
    \partial_{\pm}k^{\pm} + 2\partial_{\pm}\omega k^{\pm} &=0,
\end{align}
which is solved by
\begin{align}
    \label{eq:kexplicit}
    k^{\pm} &= C e^{-2\omega},
\end{align}
where the constant $C$ can be normalized to $1$ by an appropriate rescaling of $\lambda$.

{\it A priori} there is no notion of expansion of a unique light ray, as it is always trivially equal to zero. However, the dilaton encodes the size of the compactified space, which varies along the two-dimensional space. To compute the expansion parameter in JT gravity, we go back to its definition in terms of the extrinsic curvature, see Appendix \ref{sec:hydro_regime} and \eq{eq:defcurv}, to take into account the compact space. With the ansatz \eqref{eq:sphred}, the extrinsic curvature of the $n$-dimensional congruence of geodesics reduces to \cite{Wald:1984rg}
\begin{equation}
    {K}^{(n)} = K^{(1)} + \frac{1}{\Phi}k^{\mu}\nabla_{\mu}\Phi,
\end{equation}
where $K^{(1)}$ is the extrinsic curvature of the $1$-dimensional light ray in the two-dimensional spacetime $\mathcal{M}$. Since the extrinsic curvature of a unique light ray vanishes, we get the expression of the expansion parameter in JT gravity,
\begin{equation}
\label{eq:thJT}
    \theta = \frac{1}{\Phi}k^{\mu}\nabla_{\mu}\Phi = \frac{1}{\Phi}\frac{\d\Phi}{\d\lambda} = \frac{\d\ln \Phi}{\d\lambda}.
\end{equation}
We define a \emph{light-sheet} in JT gravity as a light ray satisfying the condition
\begin{equation}
    \Phi' = \frac{\d\Phi}{\d\lambda} \leq 0
\end{equation}
everywhere. In Appendix \ref{sec:hydro_regime}, we recall the classical focussing theorem in arbitrary dimension, which states that light rays are focused under the evolution of a null congruence if the NEC is satisfied. In order to state a two-dimensional version of the classical focussing theorem in JT gravity, we use \eqref{eq:thJT} to compute
\begin{equation}\label{eq:deriv_theta}
    \frac{\d\theta}{\d\lambda}=\frac{1}{\Phi}\frac{\d^2\Phi}{\d\lambda^2}-\theta^2.
\end{equation}
From the classical equation of motion \eqref{eq:dilaton_eom}, one sees that $\d^2\Phi/\d\lambda^2=-8\pi G~k^{\mu}k^{\nu}\braket{\tau_{\mu\nu}}$ for any null vector $k^{\mu}$, both in Anti-de Sitter and de Sitter JT gravity. This quantity is non-positive by the NEC \eqref{eq:NEC}. Since $\Phi$ is positive, $\Phi''/\Phi$ is non-positive, and we get the \emph{Classical Focussing Theorem}:
\begin{equation}
\label{eq:FT}
\frac{\d\theta}{\d\lambda}\leq 0.
\end{equation}
This inequality tells us that along a single light ray, the function $\Phi'/\Phi$ is decreasing. Therefore, if the NEC is satisfied, we can define a light-sheet as a light ray where $\Phi'\leq 0$ only initially, the classical focussing theorem then implying $\Phi'\leq 0$ all along the light ray. 

To state and prove the Bousso bound in JT gravity, one needs to define precisely what is meant by $S_{L(B-B')}$, the entropy passing through the light-sheet. For the classical Bousso bound, this is usally done using the \emph{hydrodynamic approximation}, where the matter entropy has a phenomenological description in terms of a local entropy current $s^{\mu}$. We will follow this prescription here, as well as in the next section. Alternative definitions of $S_L$ going beyond the hydrodynamic regime have been proposed in the literature, see Section \ref{sec:comparison}. The entropy flux $s$ passing through any point of the light-sheet is then the projection of $s^{\mu}$ onto the normal vector $k^{\mu}$:
\begin{equation}\label{eq:entropy_flux}
s =-k^{\mu} s_{\mu}.
\end{equation}
The total matter entropy passing through $L(B-B')$ is therefore the integral of $s$ over $L(B-B')$,
\begin{equation}
\label{eq:entL}
    S_{L(B-B')}=\int_0^1 \d \lambda~s(\lambda).
\end{equation}
The generalized covariant entropy bound in JT gravity then reads:
\begin{equation}\label{eq:2D_GCEB}
    \int_0^1 \d \lambda~s(\lambda) \leq \frac{1}{4G}\left(\Phi(0)-\Phi(1)\right),
\end{equation}
where the ``areas'' of the two endpoints of the light-sheet are given by the value of dilaton. This can be derived by dimensional reduction of the $(n+1)$-dimensional Bousso bound, as described in Appendix \ref{sec:hydro_regime}. It has been shown \cite{Bousso:2003kb,Strominger:2003br} that the classical Bousso bound can be proven in $4$ dimensions using the two conditions \eqref{eq:cond1} and \eqref{eq:cond2}. They correspond to the validity conditions for the hydrodynamic approximation, and their dimensional reduction yields
\begin{align}
    \Phi(\lambda)\left|\left(s(\lambda)\Phi^{-1}(\lambda)\right)^{'}\right| &\leq 2\pi k^{\mu}k^{\nu}\braket{\tau_{\mu\nu}}(\lambda), \label{eq:cond1JT}\\
    s(0)&\leq-\frac{1}{4G}\Phi^{'}(0) \label{eq:cond2JT}.
\end{align}
The first one implies the NEC in two dimensions, $k^{\mu}k^{\nu}\braket{\tau_{\mu\nu}}\geq 0$, while the second one implies that the light-sheet is initially non-expanding. Using the classical focusing theorem, this corresponds to imposing the non-expansion of the light ray.

\subsection{Proof of the Classical Bousso Bound}
\label{sect:proof_classical_BB}

In order to prove the classical Bousso bound in JT gravity, we first rewrite the entropy condition \eqref{eq:cond1JT} as
\begin{eqnarray}
s^{'}(\lambda)&\leq& 2\pi k^{\mu}k^{\nu}\braket{\tau_{\mu\nu}}(\lambda)+s(\lambda)\frac{\Phi^{'}(\lambda)}{\Phi(\lambda)}\\
\label{eq:cond1JTbis}
&\leq& 2\pi k^{\mu}k^{\nu}\braket{\tau_{\mu\nu}}(\lambda),
\end{eqnarray}
where the second inequality follows from $\Phi^{'}(\lambda)\leq 0$ along a light-sheet, while $s(\lambda)$ and $\Phi(\lambda)$ are both positive. On the other hand, for any null vector $k^{\mu}$, the equation of motion \eqref{eq:dilaton_eom} gives:
\begin{equation}\label{eq:eom_k_contracted}
8\pi G~ k^{\mu}k^{\nu}\braket{\tau_{\mu\nu}}=-k^{\mu}k^{\nu}\nabla_{\mu}\nabla_{\nu}\Phi=-\Phi^{''},
\end{equation}
both in Anti-de Sitter and de Sitter JT gravity, so that
\begin{equation}\label{eq:EC1_covariant_refined}
s^{'}(\lambda)\leq -\frac{1}{4G}\Phi^{''}(\lambda).
\end{equation}
From now on, we work in the conformal gauge \eqref{eq:metric_conf_gauge}, and denote by $(x^+,x^-)$ the Kruskal coordinates of an arbitrary point of the light-sheet with affine parameter $\lambda$. The initial point of the light-sheet, $\lambda=0$, is noted $(x_0^+,x_0^-)$. The derivative with respect to $\lambda$ is given by $'\equiv d/d\lambda=\frac{\partial x^{\pm}}{\partial\lambda}\partial_{\pm}$, with the $+$ or $-$ for a light-sheet along $x^+$ or $x^-$ respectively, so that the entropy conditions \eqref{eq:EC1_covariant_refined} and \eqref{eq:cond2JT} are given by:
\begin{align}
\label{eq:EC1_conf_coord}
\partial_{\pm} s ~ \frac{\partial x^{\pm}}{\partial\lambda}&\leq -\frac{1}{4G}\partial_{\pm}\left(\partial_{\pm}\Phi~\frac{\partial x^{\pm}}{\partial\lambda}\right)\frac{\partial x^{\pm}}{\partial\lambda},\\
\label{eq:EC2_conf_coord}
s(x_0^{\pm})&\leq-\frac{1}{4G}\partial_{\pm}\Phi(x_0^{\pm})~\frac{\partial x^{\pm}}{\partial\lambda}(x_0^{\pm}).
\end{align}
In the following, we will consider the case of a future directed light-sheet along $x^+$. Since $\partial x^+/\partial\lambda>0$ in this case, the first entropy condition \eqref{eq:EC1_conf_coord} gives:
\begin{equation}
\partial_+s \leq -\frac{1}{4G}\partial_+\left(\partial_+\Phi\frac{\partial x^+}{\partial\lambda}\right).
\end{equation}
Integrating this inequality between $x_0^+$ and $x^+ > x_0^+$ yields
\begin{equation}\label{eq:inequa}
s(x^+)-s(x_0^+)\leq-\frac{1}{4G}~\partial_+\Phi(x^+)~\frac{\partial x^+}{\partial\lambda}(x^+)+\frac{1}{4G}~\partial_+\Phi(x_0^+)~\frac{\partial x^+}{\partial\lambda}(x_0^+),
\end{equation}
which can be further simplified using the second entropy condition \eqref{eq:EC2_conf_coord} into:
\begin{equation}
s(x^+) \leq -\frac{1}{4G}\partial_+\Phi(x^+)~\frac{\partial x^+}{\partial\lambda}(x^+).
\end{equation}
From $s=-k^+ s_+$ and $k^+=\frac{\partial x^+}{\partial\lambda} > 0$ for a future-directed light-sheet, this gives:
\begin{equation}
-s_+(x^+)\leq -\frac{1}{4G}\partial_+\Phi(x^+).
\end{equation}
One can thus write
\begin{equation}
\int_0^1 \d\lambda~s(\lambda)=\int_{x_0^+}^{x_1^+}\d x^+\frac{\partial\lambda}{\partial x^+}(-k^+s_+) =\int_{x_0^+}^{x_1^+}\d x^+(-s_+) \leq -\frac{1}{4G}\int_{x_0^+}^{x_1^+}\d x^+\partial_+\Phi,
\end{equation}
from which we find the classical Bousso bound, written in terms of the $\lambda$ parameter:
\begin{equation}
\int_0^1 \d\lambda~s(\lambda) \leq\frac{1}{4G}(\Phi(0)-\Phi(1)).
\end{equation}
The derivation for a past-directed light-sheet along $x^+$, for which $k^+=-\frac{\partial x^+}{\partial\lambda}=e^{-2\omega}$, or for a light-sheet along $x^-$, can be carried out in a completely analogous way, leading to the same result.

\section{Strominger-Thompson Quantum Bousso Bound}
\label{sec:ST}

In this section, we prove the Strominger-Thompson Quantum Bousso Bound (QBB) in the framework of semiclassical JT gravity introduced in \Sect{sect:dS_JT}. We first consider the semiclassical violations of the classical Bousso bound and motivate the two quantum entropy conditions of \cite{Strominger:2003br}. We then introduce the quantum version of the Bousso bound that was proposed by Strominger and Thompson and proven in two specific vacua of the two-dimensional RST model \cite{Russo:1992ax}, and derive a sufficient condition for the QBB to hold in JT gravity. Inspired by the work of Wall on the generalized second law in $(1+1)$-dimensions \cite{Wall:2011kb}, we study the transformation laws of the entropy and stress tensor under a change of vacuum state, and identify a quantity invariant under conformal transformations. This is used to show that the sufficient condition is satisfied in any vacuum state defined in arbitrary lightcone coordinates, hence establishing the proof of the QBB in an infinite class of vacuum states in Anti-de Sitter and de Sitter JT gravity. 

\subsection{Generalized Entropy and Quantum Energy Conditions}
\label{sec:gen_entropy}

In our proof of the classical Bousso bound, we assumed the NEC in the assumption \eqref{eq:cond1JT}. However, this local energy condition is not a fundamental law of physics and it may be violated in the presence of matter, as well as the Bousso bound \cite{Epstein:1965zza, Lowe:1999xk, Strominger:2003br}. It is for example the case when black holes evaporate \cite{PhysRevD.13.2720}, and we expect a similar situation in (Anti-) de Sitter space. Indeed, in the JT gravity framework studied here, the total stress-energy tensor is given by $\braket{T_{\pm\pm}}=\braket{\tau_{\pm\pm}}-\frac{c}{24\pi}t_{\pm}$. While the classical part $\braket{\tau_{\pm\pm}}$ satisfies the NEC, it might be violated by the full $\braket{T_{\pm\pm}}$, depending on the functions $t_{\pm}$, which might imply a violation of the classical focussing theorem and the classical Bousso bound. This is for instance the case in the Unruh state (which is also defined in de Sitter space \cite{Aalsma:2019rpt}). One would therefore like to define a quantum version of a light-sheet, of the NEC, and of the Bousso bound that would be consistent at the semiclassical level.

The starting point for such generalizations is the observation that a quantum area (or generalized entropy) can be assigned to any point separating a $1$-dimensional Cauchy slice into two portions. The ``quantum area'' of a point $P(\lambda)$ along a light ray of affine parameter $\lambda$ is then given by
\begin{equation}\label{eq:quant_area_point}
    A_{\rm qu}(\lambda) = \Phi(\lambda) + 4G S(\lambda),
\end{equation}
where $S(\lambda)$ is the fine-grained entropy, as defined in \eq{eq:fgent}, of a Cauchy slice $\Sigma_{P}$ defined as the interior of $P(\lambda)$ \footnote{\label{fn:entropy}In a spacetime without boundary, this slice is defined such that $\partial\Sigma_P=P$. Note that there is always two choices of $\Sigma_P$ depending on what we call the interior and exterior of $P$. However, the definition of fine-grained entropy \eqref{eq:fgent} implies that $S\left(\overline{\Sigma_P}\right)=S(\Sigma_P)$, so that both choices are equivalent. Here we define $\overline{\Sigma_P}$, a complement of $\Sigma_P$, such that there exists $\Sigma$ a global Cauchy slice whose state is pure, with $\Sigma_P\cup\overline{\Sigma_P}=\Sigma$. In a closed spacetime with boundary, we define $\Sigma_P$ such that $\partial\Sigma_P=P\cup A$, where $A$ is an arbitrary point of the boundary, with the same property of interior and exterior being interchangeable.}. In analogy with the classical case, we define a \emph{quantum light-sheet} in de Sitter JT gravity as a light ray satisfying everywhere the condition
\begin{equation}\label{eq:def_quant_lightsheet}
A_{\rm qu}'(\lambda)\leq 0.
\end{equation}

At the semiclassical level, a generalisation of the NEC was proposed in \cite{Bousso:2015mna}. This so-called \emph{Quantum Null Energy Condition} (QNEC) states that:
\begin{equation}
\label{eq:QNEC}
    2\pi k^{\mu}k^{\nu}\braket{T_{\mu\nu}} \geq S''.
\end{equation}
This equation does not depend on $G$ and is unaffected by higher curvature terms in the gravitational action, such that it derives from fundamental principles in quantum field theory\footnote{Recently, an ``Improved Quantum Null Energy Condition'' has been proposed in $D \geq 4$ spacetime dimensions in \cite{Ben-Dayan:2023inz}.}, and was proven rigorously within quantum field theory in \cite{Bousso:2015wca, Balakrishnan:2017bjg}.\footnote{See\cite{Koeller:2015qmn} for an holographic proof.}. In two dimensions, it was shown in \cite{Wall:2011kb} that the QNEC implies the stronger statement
\begin{equation}
\label{eq:QNEC2D}
    2\pi k^{\mu}k^{\nu}\braket{T_{\mu\nu}} \geq S'' + \frac{c}{6}(S')^2,
\end{equation}
which follows from the transformation laws of these quantities under conformal transformation.

The Bousso bound as well as the assumptions \eqref{eq:cond1JT} and \eqref{eq:cond2JT} must be modified in order to hold at the semiclassical level. Motivated by the progresses in black hole thermodynamics and holography due to the introduction of generalized entropy \eqref{eq:sgen}, Strominger and Thompson \cite{Strominger:2003br} conjectured the \emph{Strominger-Thompson Quantum Bousso Bound}:
\begin{equation}
    \label{eq:STBB_JT}
    \int_0^1\d\lambda ~s(\lambda) \leq \frac{1}{4G}(A_{\rm qu}(0)-A_{\rm qu}(1)).
\end{equation}
Other conjectured QBB were proposed \cite{Bousso:2014sda, Bousso:2015mna} and will be discussed in Section \ref{sec:comparison}.  
At the semiclassical level, the classical stress tensor $\tau_{\mu\nu}$ must be replaced by the full stress tensor $T_{\mu\nu}$ which may violate the NEC. Since the first condition \eqref{eq:cond1JT} implies the NEC and the second one \eqref{eq:cond2JT} implies that the classical Bousso bound is satisfied at the beginning of the light-sheet, the conditions \eqref{eq:cond1JT} and \eqref{eq:cond2JT} must be modified.
Adapted to the case of JT gravity, the semiclassical conditions in two dimensions are:
\begin{itemize}
    \item The first classical condition \eqref{eq:cond1JTbis} is unchanged,
    \begin{equation}
    \label{eq:cond1_QNEC}
    \left|s^{'}(\lambda)\right|\leq 2\pi k^{\mu}k^{\nu}\braket{\tau_{\mu\nu}}(\lambda),
    \end{equation}
    without introducing the quantum mechanical part of the stress tensor, which corresponds to imposing the classical condition in the non-backreacted geometry. It is well defined since $\tau_{\mu\nu}$ always satisfies the NEC.
    \item The modification $A\rightarrow A_{\rm qu}$ is applied to the second classical condition \eqref{eq:cond2JT}, implying that the QBB is initially satisfied:
    \begin{equation}
    \label{eq:cond2_QNEC}
        s(0) \leq -\frac{1}{4G}A_{\rm qu}'(0).
    \end{equation}
    Since $s(\lambda)$ is positive, this condition implies the quantum area to be initially non-increasing. In particular, this is satisfied by a quantum light-sheet, as defined in \eqref{eq:def_quant_lightsheet}.
\end{itemize}

\subsection{A sufficient condition}
\label{sect:quantum_bound}

In order to investigate the QBB in (Anti-)de Sitter JT gravity, we start from the equation of motion at the semiclassical level \eqref{eq:dilaton_eom}. For any null vector $k^{\pm}$, this gives:
\begin{equation}
8\pi G k^{\pm} k^{\pm}\left(\braket{\tau_{\pm\pm}}-\frac{c}{24\pi}t_{\pm}\right)=-\Phi''=-\partial_{\pm}\left(\partial_{\pm}\Phi\frac{\partial x^{\pm}}{\partial\lambda}\right)\frac{\partial x^{\pm}}{\partial\lambda},
\end{equation}
independently of the sign of the cosmological constant $\Lambda$. Inserting this equation into the first entropy condition \eqref{eq:cond1_QNEC} leads to the inequality:
\begin{equation}
\partial_{\pm}s~\frac{\partial x^{\pm}}{\partial\lambda}\leq -\frac{\Phi''}{4G}+\frac{c}{12}k^{\pm} k^{\pm} t_{\pm},
\end{equation}
or equivalently, using the definition of the quantum area of a point \eqref{eq:quant_area_point}:
\begin{equation}
\partial_{\pm} s~\frac{\partial x^{\pm}}{\partial\lambda}\leq -\frac{A_{\rm qu}''}{4G}+S''+\frac{c}{12}k^{\pm}k^{\pm} t_{\pm}.
\end{equation}
As it can be seen from this last inequality, a sufficient condition for the QBB \eqref{eq:STBB_JT} to be satisfied is
\begin{equation}\label{eq:sufficient_cond}
S''+\frac{c}{12}k^{\pm} k^{\pm} t_{\pm}\leq 0.
\end{equation}
Written in terms of the normal ordered stress tensor, this condition is
\begin{equation}
    2\pi k^{\pm}k^{\pm}\braket{:T_{\pm\pm}:}\geq S'',
\end{equation}
which closely resembles the QNEC \eqref{eq:QNEC}. In particular, it becomes exactly the QNEC in the limit $\tau_{\pm\pm}=0$. If the inequality \eqref{eq:sufficient_cond} is true, one gets 
\begin{equation}\label{eq:starting_inequality}
\partial_{\pm}s~\frac{\partial x^{\pm}}{\partial\lambda}\leq -\frac{A_{\rm qu}''}{4G},
\end{equation}
from which one can follow the same steps as in the proof of the classical Bousso bound presented in \Sect{sect:proof_classical_BB}. 

Here, we consider as an example the case of a past directed light-sheet along $x^+$. Since $\partial x^+/\partial\lambda<0$, \eqref{eq:starting_inequality} gives:
\begin{equation}
\partial_+s \geq -\frac{1}{4G}\partial_+\left(\partial_+A_{\rm qu}\frac{\partial x^+}{\partial\lambda}\right).
\end{equation}
Integrating this inequality between $x^+$ and $x_0^+>x^+$ yields
\begin{equation}
s(x_0^+)-s(x^+)\geq-\frac{1}{4G}~\partial_+A_{\rm qu}(x_0^+)~\frac{\partial x^+}{\partial\lambda}(x_0^+)+\frac{1}{4G}~\partial_+A_{\rm qu}(x^+)~\frac{\partial x^+}{\partial\lambda}(x^+),
\end{equation}
which can be further simplified using the second entropy condition \eqref{eq:cond2_QNEC} into
\begin{equation}
-s(x^+) \geq \frac{1}{4G}\partial_+A_{\rm qu}(x^+)~\frac{\partial x^+}{\partial\lambda}(x^+).
\end{equation}
From $s=-k^+ s_+$ and $k^+=-\frac{\partial x^+}{\partial\lambda} > 0$ for a past-directed light-sheet, this gives:
\begin{equation}
-s_+(x^+)\leq \frac{1}{4G}\partial_+A_{\rm qu}(x^+).
\end{equation}
One can thus write
\begin{equation}
\int_0^1 \d\lambda~s(\lambda)=\int_{x_0^+}^{x_1^+}\d x^+\frac{\partial\lambda}{\partial x^+}(-k^+s_+) =\int_{x_1^+}^{x_0^+}\d x^+(-s_+) \leq \frac{1}{4G}\int_{x_1^+}^{x_0^+}\d x^+\partial_+A_{\rm qu},
\end{equation}
from which we find the QBB:
\begin{equation}
\int_0^1 \d\lambda~s(\lambda)\leq\frac{1}{4G}(A_{\rm qu}(0)-A_{\rm qu}(1)).
\end{equation}
We thus showed that the inequality \eqref{eq:sufficient_cond} is a sufficient condition for the Strominger-Thompson QBB to hold in the (Anti-)de Sitter JT gravity framework.

\subsection{Proof of the Quantum Bousso Bound}
\label{sec:sufcondproof}

To show that the sufficient condition \eqref{eq:sufficient_cond} is verified in any vacuum state, we consider the transformation laws of $S$ and $t_{\pm}$ under a change of vacuum, \ie under a conformal transformation of the lightcone coordinates in which the vacuum is defined. 
Under a change of coordinates $x^{\pm}\rightarrow y^{\pm}(x^{\pm})$, the metric becomes
\begin{align}
\d s^2 &= -e^{2\omega_x(x^+,x^-)}\d x^+\d x^- = -e^{2\omega_x(x^+,x^-)}\frac{\d x^+}{\d y^+}\frac{\d x^-}{\d y^-}\d y^+\d y^-\\
&= -e^{2\omega_y(y^+,y^-)}\d y^+\d y^-,
\end{align}
from which we get the transformation of the conformal factor:
\begin{equation}
\label{eq:transfo_conf_factor}
    \omega_{y}(y^+(x^+),y^-(x^-)) = \omega_{x}(x^+,x^-) -\frac{1}{2}\ln\left[\frac{\d y^+}{\d x^+}\frac{\d y^-}{\d x^-}\right].
\end{equation}
We start by explicitly study the transformation law of the entropy $S(\Sigma)$ in the half reduction model of de Sitter JT gravity, given in \eqref{eq:ent_bdy_BD} or \eqref{eq:ent_bdy}. As will be discussed below, the analysis in the full reduction model of de Sitter JT gravity, as well as in Anti-de Sitter JT gravity, is analogous and leads to the same result. To make the dependence in the boundaries of the half reduction model clear, we keep the general formulation \eqref{eq:ent_bdy} where the boundary follows a trajectory $x^-=f_x(x^+)$, as described in \Appendix{app:reflec_bound}. The transformation law of the entropy under a change of vacuum $x^{\pm}\rightarrow y^{\pm}(x^{\pm})$ is (see \eq{eq:transf_bdy}),
\begin{align}
    S_x(\Sigma)\rightarrow S_y(\Sigma)
    &= \frac{c}{6}\omega_x(x^+,x^-) + \frac{c}{12}\ln\left[\dot f_x(f_x^{-1}(x^-))\frac{\left(y^+(x^+) -y^+(f_x^{-1}(x^-))\right)^2}{\delta^2}\right]\nonumber\\
    &-\frac{c}{12}\ln\left[\frac{\d y^+}{\d x^+}(x^+)\frac{\d y^+}{\d x^+}\left(f_x^{-1}(x^-)\right)\right],
\end{align}
where $\dot f_x = \d f_x/\d x^+$. From this relation, one can then compute the transformation laws for the first and second derivatives of the entropy. We take the case were the derivation is taken along a light-sheet going in the $x^+$ direction, \ie $'\equiv\d/\d\lambda = k^+\partial/\partial x^+$, giving
\begin{equation}
S_x'(\Sigma)\rightarrow S_y'(\Sigma)=\frac{c}{6}\left[\frac{\partial \omega_x}{\partial x^+}-\frac{1}{2}\frac{\ddot y^+}{\dot y^+}+\frac{\dot y^+}{y^+-f_y^{-1}(y^-)}\right]\frac{\partial x^+}{\partial \lambda},
\end{equation}
where $\dot y^+ = \d y^+/\d x^+$. Taking the second derivative, we get
\begin{eqnarray}
S_x''(\Sigma)&\rightarrow& S_y''(\Sigma)=\frac{c}{6}\left[\frac{\partial^2 \omega_x}{\partial x^{+2}}-\frac{1}{2}\frac{\dddot y^+}{\dot y^+}+\frac{1}{2}\left(\frac{\ddot y^+}{\dot y^+}\right)^2-\frac{(\dot y^+)^2}{(y^+-f_y^{-1}(y^-))^2}\right.\nonumber\\
&+&\!\!\!\!\!\left.\frac{\ddot y^+}{y^+-f_y^{-1}(y^-)}-2\frac{\partial\omega_x}{\partial x^+}\left(\frac{\partial\omega_x}{\partial x^+}-\frac{1}{2}\frac{\ddot y^+}{\dot y^+}+\frac{\dot y^+}{y^+-f_y^{-1}(y^-)}\right)\right]\left(\frac{\partial x^+}{\partial \lambda}\right)^2.
\end{eqnarray}
Combining the two expressions above, we obtain the transformation law for the quantity $S''+\frac{6}{c}(S')^2$ introduced by Wall \cite{Wall:2011kb}:
\begin{equation}
S_x''+\frac{6}{c}(S_x')^2\rightarrow S_y''+\frac{6}{c}(S_y')^2 = \frac{c}{6}\left[\frac{\partial^2\omega_x}{\partial x^{+2}}-\left(\frac{\partial\omega_x}{\partial x^+}\right)^2-\frac{1}{2}\{y^+,x^+\}\right]\left(\frac{\partial x^+}{\partial \lambda}\right)^2, 
\end{equation}
which can be rewritten as
\begin{equation}\label{eq:trans_law_S''}
S_x''+\frac{6}{c}(S_x')^2\rightarrow S_y''+\frac{6}{c}(S_y')^2 = S_x''+\frac{6}{c}(S_x')^2-\frac{c}{12}\{y^+,x^+\}\left(\frac{\partial x^+}{\partial \lambda}\right)^2.
\end{equation}
Similarly, we can compute the $t_{\pm}(x^{\pm})$ function in the vacuum defined by $t_{\pm}(y^{\pm})=0$ using \eq{eq:transft}. As for the entropy, we will make the vacuum state explicit by writing $t^{(y)}_{\pm}$ the $t_{\pm}$ function in the vacuum defined in coordinates $y^{\pm}$, \ie $t^{(y)}_{\pm}(y^{\pm})=0$. The transformation under a change of vacuum then writes
\begin{equation}
   t^{(x)}_{\pm}(x^{\pm}) = 0 \rightarrow t^{(y)}_{\pm}(x^{\pm}) = \left\{y^{\pm},x^{\pm}\right\}.
\end{equation}
From the transformation laws of $S$, its derivatives, and $t_{+}$, we find that the quantity 
\begin{equation}
    \mathcal{Q} = \frac{c}{12} k^{+}k^{+}t^{(x)}_{+} + S_x'' + \frac{6}{c}(S_x')^2
\end{equation}
is a scalar under conformal transformation. The choice of initial vacuum state being arbitrary, we can drop the vacuum indices and write
\begin{equation}
    \mathcal{Q} = \frac{c}{12} k^{+}k^{+}t_{+} + S'' + \frac{6}{c}(S')^2.
\end{equation}
The argument above can be extended to a light-sheet propagating along the $x^-$ direction. This may seem cumbersome, due to the dependence of $\dot f_x(f_x^{-1}(x^-))$ in $x^-$. However, the derivation of \eqref{eq:ent_bdy} in Appendix \ref{app:entanglement_entropy} can be modified to obtain a very similar formula depending on $\dot f_x(f_x(x^+))$\footnote{To do so, one describes the state of fields as right-moving modes on $H_-$ instead of left-moving modes on $H_+$, see Appendix \ref{app:entanglement_entropy}.}, allowing to follow the same procedure for a light-sheet along $x^-$.

At this point, we can notice that in the Bunch-Davies vacuum, $t_{\pm}(x^{\pm})=0$ and
\begin{equation}
\label{eq:BDeq}
    S_{\rm BD}'' + \frac{6}{c}(S_{\rm BD}')^2=0,
\end{equation}
as follows from \eq{eq:entang_entropy}, so that $\mathcal{Q}=0$. Because $\mathcal{Q}$ is a scalar under conformal transformations, this must be true in any vacuum state. To conclude, we have shown that
\begin{equation}
    S''+\frac{c}{12} k^{\pm}k^{\pm}t_{\pm}  = - \frac{6}{c}(S')^2 \leq 0
\end{equation}
in any vacuum state, hence completing the proof of the Strominger-Thompson QBB. This inequality together with the NEC satisfied by the classical part $\braket{\tau_{\pm\pm}}$ of the stress tensor implies the QNEC. In particular, it implies the stronger inequality \eqref{eq:QNEC2D}, showed to be equivalent to the QNEC by Wall in two dimensions \cite{Wall:2011kb}. 

\begin{figure}[h!]
\centering
\begin{tikzpicture}
\begin{scope}[transparency group]
\begin{scope}[blend mode=multiply]
\path
       +(6,3)  coordinate (IItopright)
       +(0,3)  coordinate (IItopcenter)
       +(-6,3) coordinate (IItopleft)
       +(0,3)  coordinate (IIbotcenter)
       +(6,-3) coordinate (IIbotright)
       +(-6,-3) coordinate(IIbotleft)
      
       ;

\draw (-3,3) -- (3,3) ;
\draw (-3,-3) -- (3,-3) ;
\draw (-6,-2) -- (-6,2) ;
\draw (6,-2) -- (6,2) ;

\draw (-3,3) -- (-6,0) -- (-3,-3) ;
\draw (3,3) -- (6,0) -- (3,-3) ;
\draw (-3,3) -- (3,-3) ;
\draw (-3,-3) -- (3,3) ;

\fill[fill=gray!20] (-6,0) -- (-3,3) -- (0,0) -- (-3,-3);
\fill[fill=gray!20] (0,0) -- (3,-3) -- (-3,-3) -- cycle;

\fill[fill=gray!20] (-3,3) decorate[decoration=snake] {to[bend left=10] (-6,2)} -- (-6,0);
\fill[fill=gray!20] (3,3) decorate[decoration=snake] {to[bend left=-10] (6,2)} -- (6,0);
\fill[fill=gray!20] (-3,-3) decorate[decoration=snake] {to[bend left=-10] (-6,-2)} -- (-6,0);
\fill[fill=gray!20] (3,-3) decorate[decoration=snake] {to[bend left=10] (6,-2)} -- (6,0);

\draw[dotted] (-3,-3)--(-3,3);
\draw[dotted] (3,3)--(3,-3);

\draw[decorate,decoration=snake] (-3,3) to[bend left=10] (-6,2);
\draw[decorate,decoration=snake] (-3,-3) to[bend left=-10] (-6,-2);
\draw[decorate,decoration=snake] (3,3) to[bend left=-10] (6,2);
\draw[decorate,decoration=snake] (3,-3) to[bend left=10] (6,-2);

\draw (2,1) to[bend left=10] node[midway, below, sloped] {{\footnotesize $\Sigma_P$}}(6,0);

\node at (2,1) [label = above:{$P$}]{};
\node at (2,1) [circle, fill, inner sep=1.5 pt]{};

\node at (4,-1) [label = left:{$B'$}]{};
\node at (4,-1) [circle, fill, inner sep=1.5 pt]{};

\node at (1,2) [label = left:{$B$}]{};
\node at (1,2) [circle, fill, inner sep=1.5 pt]{};

\node at (6,0) [label = right:{$A$}]{};
\node at (6,0) [circle, fill, inner sep=1.5 pt]{};

\draw[dashed] (4,-1)--
(1,2);

\draw [-{Stealth[length=3mm]},blue,decorate,decoration={snake, post length=2mm}](2,-2) -- (3,-1) ;
\draw [-{Stealth[length=3mm]},blue,decorate,decoration={snake, post length=2mm}](1,-1) -- (2,0);

\end{scope}
\end{scope}
\end{tikzpicture}
    \caption{\footnotesize Penrose diagram for two-dimensional de Sitter space in the full reduction model. A portion of a light ray bounded by two points $B$ and $B'$ is depicted by the dashed line, as well as a Cauchy slice $\Sigma_P$ bounded by a point $P$ of the light ray and the point $A$ at spatial infinity. In the Unruh-de Sitter vacuum, the dilaton diverges on the past cosmological horizon and on the black hole horizon of the static patch of the antipode, eliminating the gray shaded regions from the backreacted solution. The past horizon emits radiation (in blue) crossing $\Sigma_P$ as right-moving modes, and ending up in the weakly gravitating region at $x^+\rightarrow\infty$.}
    \label{fig:Penrose_diag_full_LS}
\end{figure}
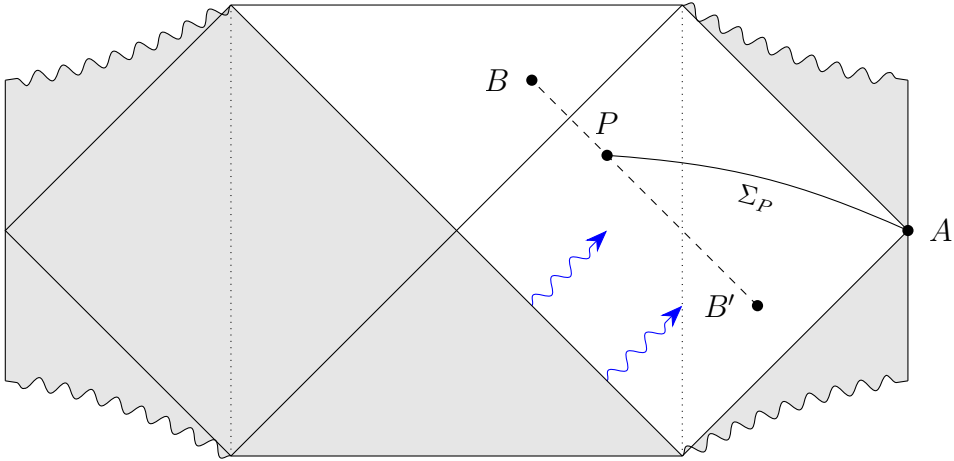

A completely similar analysis can be carried out in the full reduction model of de Sitter JT gravity, as well as in Anti-de Sitter JT gravity, considering now the standard entropy formula \eqref{eq:usualentropy_x}. Let us note that in the full reduction model of de Sitter JT, there is \apriori an ambiguity in the definition of the slice $\Sigma_P$ associated to a point $P$. Without matter, the full reduction model is either spatially periodic or infinitely extended, and one would need at least two points to define the boundary of a non-trivial Cauchy slice. Hence, we will always consider one of these two points to be fixed and spacelike separated from the second one moving along the light-sheet. 
This produces the same transformation laws as in the half reduction model, so that $\mathcal{Q}$ is invariant. In particular, $S''+ 6/c (S')^2$ does not depend on the position of the fixed point. Using the fact that $t_{\pm}^{(x)}(x^{\pm})=0$ in the Bunch-Davies/Hartle-Hawking states, and that the entropy formula \eqref{eq:usualentropy_x} also satisfies \eqref{eq:BDeq}, one recovers the result that $\mathcal{Q}=0$ in any vacuum state. This situation is depicted for the dS full reduction model in the Penrose diagram of \Fig{fig:Penrose_diag_full_LS}, in the case where the fixed point $A$ lies at $x_A^{\pm}=\pm\infty$. This is a preferred point since it is the only one spacelike separated to any light-sheet contained in the white and/or blue regions of \Fig{fig:Penrose_diag_full}. One may also be tempted to define the problem such that both endpoints of the slice belong to two lightrays forming a disconnected light-sheet. We will not treat this problem in this paper although we expect it is well defined in de Sitter JT gravity. 

Finally, we may note that the backreaction of matter on the dilaton modifies the effective geometry, which can remove the ambiguity due to spatial periodicity/infinite extension in some vacuum states. For example, the dilaton may diverge in some regions of the full reduction model, creating effective boundaries. This is the case in the Unruh-de Sitter vacuum \cite{Aalsma:2019rpt, Kames-King:2021etp, Aalsma:2021bit}, in which the backreacted geometry ends on the past cosmological horizon and black hole horizons of one of the static patches\cite{Aalsma:2021bit},  see Figure \ref{fig:Penrose_diag_full_LS}. This vacuum state describes the evaporation of the de Sitter cosmological horizon, and is analogous to the Unruh vacuum of a black hole. In this example, there is a conformal boundary at $x^+\rightarrow +\infty$, which acts as a weakly gravitating region where the radiation emanating from the past cosmological horizon can be collected. In this type of situation, one places the fixed point $A$ at the effective boundary as in Figure \ref{fig:Penrose_diag_full_LS}.

\subsection{Comparison to other Quantum Bousso Bounds}
\label{sec:comparison}

Following the proposal of Strominger and Thompson, other conjectures were made to include semiclassical corrections in the Bousso bound. Instead of applying the transformation $A\rightarrow A_{\rm qu}$, these Quantum Bousso Bounds modify the definition of $S_{L(B-B')}$ to go beyond the hydrodynamic regime, while keeping the right-hand side of the classical bound unchanged. The most recent one was obtained by Bousso, Fisher, Leichenauer, and Wall (BFLW) as a corollary of the Quantum Focussing Conjecture (QFC) \cite{Bousso:2015mna}, a semiclassical extension of the classical focussing theorem.\footnote{This bound should not be confused with the QNEC, although they were conjectured in the same paper. In particular, it contains Newton's constant and is not a purely QFT statement.} This QBB states that the entropy of the light-sheet $L(B'-B)$ can be defined as the difference between the fine-grained entropies of Cauchy slices bounded by the codimension $2$ surfaces $B$ and $B'$ respectively, and applies to quantum light-sheets instead of classical light-sheets. On the other hand, Bousso, Casini, Fisher, and Maldacena (BCFM) \cite{Bousso:2014sda} showed that in the weak gravity limit, $S_{L(B-B')}$ may be defined as a difference of fine-grained entropies of an arbitrary state and the vacuum, with both states restricted to $L(B-B')$. In this section, we review how the BFLW quantum Bousso bound can be derived from the QFC and note that it follows from our proof of the Strominger-Thompson bound. Finally, we recall the BCFM quantum Bousso bound and compare it with the two others.

\subsubsection{BFLW Quantum Bousso Bound from the QFC}

In two dimensions, a quantum expansion $\Theta$ can be defined by:
\begin{equation}
    \Theta =  \frac{1}{\Phi (\lambda)}\frac{\d}{\d\lambda} A_{\rm qu}(\lambda),
\end{equation}
which reduces to the classical expansion \eqref{eq:thJT} in the classical limit.\footnote{Reinserting factors of $\hbar$ in the quantum area gives $A_{\rm qu} = \Phi + 4G \hbar S$, hence yielding the classical result $\Theta\rightarrow\theta$ in the limit $\hbar\rightarrow 0$.} The \emph{Quantum Focussing Conjecture} in two dimensions then states that the quantum expansion cannot increase along any light ray \cite{Bousso:2015mna}:
\begin{equation}
\label{eq:QFCJT}
   \frac{\d\Theta}{\d\lambda}\leq 0.
\end{equation}
Since $\Phi>0$, if $A_{\rm qu}'$ is initially non-positive, the QFC implies that $A_{\rm qu}'(\lambda)\leq 0$ all along the light ray. If the QFC is satisfied, the definition \eqref{eq:def_quant_lightsheet} of a quantum light-sheet can therefore be restricted to $A_{\rm qu}'(0)\leq 0$ only initially, the QFC then implying that $A_{\rm qu}'(\lambda)\leq 0$ at all later points $\lambda\geq 0$ of the light ray.

As a supporting evidence for the QFC, it was noted in \cite{Bousso:2015mna} that it implies a QBB. Indeed, the property $A_{\rm qu}'(\lambda)\leq 0$, following from $A_{\rm qu}'(0)\leq 0$ and the QFC, gives in particular 
\begin{equation}
\label{eq:BFLWbound}
    0\leq \frac{1}{4G}(A_{\rm qu}(0)-A_{\rm qu}(1)),
\end{equation}
which is cutoff-independent. This can be seen as a quantum version of the Bousso bound \eqref{eq:2D_GCEB}, where $S_{L(B'-B)}$ has been identified with $S(1)-S(0)$:
\begin{equation}
    S(1)-S(0)\leq \frac{1}{4G}(\Phi(0)-\Phi(1)).
\end{equation}
This bound and the Strominger-Thompson bound have similar forms. Though, they are not equivalent since the Strominger-Thompson bound is a stronger inequality relying on stronger conditions valid only in the hydrodynamic limit, which breaks down at small scales. 

In fact, formally removing the coarse-grained entropy flux from the conditions \eqref{eq:cond1_QNEC} and \eqref{eq:cond2_QNEC} gives
\begin{align}
    k^{\mu}k^{\nu}\braket{\tau_{\mu\nu}}&\geq 0,\\
    A_{\rm qu}'(0)&\leq 0,
\end{align}
from which one can follow through our proof and recover the BFLW quantum Bousso bound, without relying on the hydrodynamic limit nor on the QFC.

\subsubsection{BCFM Quantum Bousso Bound}

Similarly to the BFLW quantum Bousso bound, the BCFM quantum Bousso bound \cite{Bousso:2014sda} proposes a modification of the left-hand side of \eqref{eq:2D_GCEB} in order to define a cutoff-independent quantity. In their work, the entropy of the light-sheet $S_{L(B'-B)}$ is defined as the difference $\Delta S$ between the fine-grained entropies of the state under consideration and the vacuum, restricted to $L(B'-B)$. A proper definition of $\Delta S$ may be found in \cite{Casini:2008cr,Bousso:2014sda}. Moreover, the BCFM bound applies to light-sheets in their classical sense, contrary to the Strominger-Thompson and BFLW bounds that apply to quantum light-sheets. The quantity $\Delta S$ is divergence-free and the QBB is proven for any portion of the light-sheet. In particular, $\Delta S$ is well defined for light-sheets of arbitrary sizes, whereas the hydrodynamic limit used in the Strominger-Thompson bound breaks down at small scales. 

This definition of entropy on the light-sheet is only valid in the weak gravity limit, \ie when the spacetime geometry in the presence of matter is well approximated by the vacuum geometry. Outside this limit, the backreaction of matter on the spacetime geometry is non-negligible and $\Delta S$ is not well defined. Indeed, the meaning of ``same'' light-sheets in two different geometries is unclear. We do not consider this version of the bound in this work. Further studies of the BCFM bound may be interesting since it is expected to not be equivalent to the BFLW bound \cite{Bousso:2015mna}, as $\Delta S$ is intrinsically local to the light-sheet while the BFLW bound is fundamentally non-local and relies on a different notion of light-sheet (quantum light-sheet).

\section{Conclusion}
\label{sec:_conclusion}

In this work, we have proven the Strominger-Thompson quantum Bousso bound in the infinite class of conformal vacua in Anti-de Sitter and de Sitter JT gravity coupled to a CFT with central charge $c$. The proof relies on a general sufficient condition, which is checked by deriving a stronger version of the QNEC. The BFLW quantum Bousso bound follows directly from our argument, without assuming the QFC. Here we comment briefly on possible future works.


As mentioned above, we do not prove the Quantum Focussing Conjecture in this work. The verification of the BFLW quantum Bousso bound provides good evidence that the QFC should be valid in the models studied here, but one would be interested in an explicit proof of this fundamental property.

Finally, we derived an entropy formula in the presence of reflective boundaries. In the presence of these boundaries, left and right-moving modes are correlated. As a consequence, the entropy of a Cauchy slice bounded by a point and the boundary cannot depend on the endpoint at the boundary. It would be interesting to consider the problem of islands and information recovery in the half reduction model of de Sitter JT gravity using this formula. It may also be used in an attempt to compute holographic entanglement entropies in the framework of de Sitter holography.

\section*{Acknowledgements}
We are grateful to Alex Kehagias, Andrew Svesko, and Watse Sybesma for interesting discussions during early stages of this research. We thank Hervé Partouche and Nicolaos Toumbas for helpful discussions and feedback. 
F.R. would like to acknowledge the hospitality of the Ecole Polytechnique, while V.F. would like to thank the University of Cyprus for hospitality, where early stages of this work have been done. This work is partially supported by the Cyprus Research and Innovation Foundation grant EXCELLENCE/0421/0362.


\appendix
\numberwithin{equation}{section}

\section{Entanglement entropy and reflecting boundaries}
\label{app:entanglement_entropy}

In this Appendix, we aim to provide a general entanglement entropy formula in a two-dimensional curved background, with two reflective boundaries. We assume massless scalar fields and follow closely the prescription of \cite{Fiola:1994ir}. 

Considering a Cauchy slice $\Sigma_{\mathcal{M}}$ of the two-dimensional spacetime $\mathcal{M}$ described by a state $\ket{\Psi}$, the state of the fields in a subsystem $\Sigma \subset \Sigma_{\mathcal{M}}$ is described by the density matrix
\begin{equation}
    \rho = \tr_{\bar{\Sigma}}{\ket{\Psi}\bra{\Psi}},
\end{equation}
where $\tr_{\bar{\Sigma}}$ traces out the degrees of freedom on $\bar{\Sigma}=\Sigma_{\mathcal{M}}\backslash \Sigma$. The entanglement entropy (or fine-grained entropy) of $\Sigma$ is defined as
\begin{equation}
\label{eq:fgent}
    S(\Sigma) = - \tr_{\Sigma} \rho\ln\rho,
\end{equation}
where $\tr_{\Sigma}$ traces over the degrees of freedom on $\Sigma$.

\subsection{Entanglement entropy without boundaries}

Let us consider a Cauchy slice $\Sigma$ in a curved two-dimensional spacetime with metric \eqref{eq:metric_conf_gauge} and without boundaries, and review some results of \cite{Holzhey:1994we,Fiola:1994ir}. For free massless scalars, the right-moving and left-moving modes are uncoupled such that we can consider them separately. We consider left-moving modes and note the coordinates of the endpoints of $\Sigma$ as $(x_1^+,x_1^-)$ and $(x_2^+,x_2^-)$, such that left-moving modes along the interval $[x_1^+,x_2^+]$ all pass through $\Sigma$. By constructing a complete set of left-moving modes localized inside and outside of the interval, one can then trace on the vacuum state built on the $x^{\pm}$ coordinates. The entropy has logarithmic UV divergences at each endpoint of the interval, arising from short wavelength fluctuations around these endpoints. They can be regulated by excluding contributions within a short distance $\varepsilon_{\Le}(x_1)$ and $\varepsilon_{\Le}(x_2)$ around $x_1^+$ and $x_2^+$, respectively. Similarly, an infrared cutoff $L$ is introduced, such that $x^+$ is restricted to the range $[-L,L]$. When the size of the interval $|x_2^+-x_1^+|$ is small enough compared to the infrared cutoff, we get
\begin{equation}
    \label{eq:entropy_left}
    S(\Sigma) = \frac{c}{12}\ln\left[\frac{(x_2^+-x_1^+)^2}{\varepsilon_{\Le}(x_1)\varepsilon_{\Le}(x_2)}\right].
\end{equation}
The cutoffs $\varepsilon_{\Le(\Ri)}$ are expressed in terms of the $x^{\pm}$ coordinates. One can rewrite these cutoffs in terms of cutoffs $\delta_{\Le(\Ri)}$ in inertial coordinates according to
\begin{equation}
\label{eq:cutoff}
    \varepsilon_{\Le(\Ri)}(x^+,x^-) = e^{-\omega(x^+,x^-)}~ \delta_{\Le(\Ri)},
\end{equation}
so that the product of the cutoffs for left and right-moving modes is boost-invariant. Adding the contribution from the right-moving modes leads to the total entropy of $\Sigma$,
\begin{equation}
    S(\Sigma) = \frac{c}{6}\left(\omega(x_1)+\omega(x_2)\right) + \frac{c}{12}\ln\left[\frac{(x_2^+-x_1^+)^2(x_2^--x_1^-)^2}{\delta_1^2\delta_2^2}\right],
\end{equation}
where $\delta_{i}^2=\delta_{\Le}(x_i)\delta_{\Ri}(x_i)$, for $i=1,2$. In two-dimensional space, $\delta_i^2$ can be seen as a numerical constant \cite{Fiola:1994ir}.

In such spacetime without boundaries, the right and left-moving modes are completely independent. But in a model with reflecting boundaries like the half reduction model of de Sitter JT gravity, the reflection of right-moving modes into left-moving modes and vice-versa induces correlations between them. A study of entanglement entropy in the presence of a single moving mirror following the timelike trajectory $x^+ - x^- =$ constant has been provided in \cite{Fiola:1994ir}. In the next section, we discuss a more general case where the spacetime contains two reflecting boundaries following more general trajectories.

\subsection{Entanglement entropy with reflecting boundaries}
\label{app:reflec_bound}

Let us consider the Kruskal coordinates $(x^+,x^-)$ and the metric \eqref{eq:metric_conf_gauge} in the conformal gauge. Such set of coordinates naturally splits the spacetime into four distinct regions. Inspired by the half reduction model of de Sitter JT gravity, we consider the case where the spacetime contains two spatial reflecting boundaries lying entirely in the regions $x^+ x^-\leq 0$. We parameterize their trajectories through a function\footnote{One could find a change of coordinates $y^{\pm}(x^{\pm})$ such that the boundary is described by $y^+ - y^- =$ constant. However, choosing a system of coordinates in \eqref{eq:entropy_left} corresponds to selecting a specific vacuum, defined in these coordinates. It is therefore convenient to have the freedom to choose the coordinate system in which we define the vacuum.} $x^-=f(x^+)$, hence satisfying $x^+ f(x^+)\leq 0$, and impose in addition the following conditions:
\begin{enumerate}
    \item Introducing an infrared cutoff $L$, the left boundary is bounded by the points of $x^{\pm}$ coordinates $(-L,0)$ and $(0,L)$, while the right boundary is bounded by the points $(0,-L)$ and $(L,0)$. We thus have $f:I \rightarrow I$, with  $I=(-L,0)\cup (0,L)$.
    \item All modes emanating from any Cauchy slice bounded by the points $(-L,0)$ and $(0,-L)$ reflect once and only once on a boundary, which is equivalent to have $f$ a bijection from $I$ to $I$.
    \item The trajectory must be continuous and differentiable, which is equivalent to impose $f$ to be a $C^1$ function.
    \item The trajectory must be timelike to define a spatial boundary, which is equivalent to have $0<\dot f<+ \infty$, where $\dot f\equiv \d f/\d x^+$.
    \item We restrict ourselves to states that have symmetric left and right-moving radiation, and in general to systems symmetric under $x^+\leftrightarrow x^-$. This implies that $f$ is an involution, \ie $f=f^{-1}$.
\end{enumerate}
This construction ensures that, when a mode reflect on a boundary:
\begin{itemize}
    \item Left-moving modes propagating along $x^-$ at fixed $x^+=x_0^+$ transform into right-moving modes propagating along $x^+$ at fixed $x^-=f(x_0^+)$.
    \item Right-moving modes propagating along $x^+$ at fixed $x^-=x_0^-$ transform into left-moving modes propagating along $x^-$ at fixed $x^+=f(x_0^-)$.
\end{itemize}
In particular, we are able to express all modes either as right-moving modes emanating from the interval $[-L,L]$ on the $x^-$ axis, or as left-moving modes emanating from the interval $[-L,L]$ on the $x^+$ axis.

\begin{figure}[h!]
\begin{subfigure}[t]{0.48\linewidth}
\centering
\begin{tikzpicture}
\path
       +(3,3)  coordinate (IItopright)
       +(-3,3) coordinate (IItopleft)
       +(3,-3) coordinate (IIbotright)
       +(-3,-3) coordinate(IIbotleft)
      
       ;
\draw (IItopleft) --
          node[midway, above, sloped] {{\footnotesize$x^-=g(x^+)$}}
      (IItopright) --
          node[midway, above,sloped] {{\footnotesize $x^-=f(x^+)$}}
      (IIbotright) -- 
          node[midway, below, sloped] {}
      (IIbotleft) --
          node[midway, above , sloped] {{\footnotesize $x^-=f(x^+)$}}
      (IItopleft) -- cycle;

\draw (IItopleft) -- (0,0) -- (IItopright);
\draw (IIbotleft) -- (-2.2,-2.2);
\draw[blue] (-2.2,-2.2) -- (-3.2/2,-3.2/2);
\draw[red] (-3.2/2,-3.2/2) -- (-1/2,-1/2);
\draw (-1/2,-1/2) -- (0,0) -- (IIbotright);

\draw (IItopleft) -- (-3.5,3.5) ;
\draw (IItopright) -- (3.5,3.5) ;
\draw (IItopright) -- (3.5,3.5) ;
\draw (-3.5,3.3) -- (-3.5,3.5) -- node[midway, above, sloped] {$x^-$} (-3.3,3.5) ;
\draw (3.5,3.3) -- (3.5,3.5) -- node[midway, above, sloped] {$x^+$} (3.3,3.5) ;

\draw (-3,-0.2) to [bend right=15] (-1.3,0.3);
\draw[dotted] (-1.3,0.3) -- (-1/2,-1/2);
\draw[dotted] (-3,-0.2) -- (-3.2/2,-3.2/2);
\draw[dotted] (-1.3,0.3) -- (-3,-3+1.3+0.3);
\draw[dotted] (-0.8,0.8) -- (-1.3,0.3);
\draw[dotted] (-3,-3+1.3+0.3) -- (-2.2,-2.2);

\node at (-0.7,-0.7) [label = right:{\footnotesize $x^+$}]{};
\node at (-1.8,-1.8) [label = right:{\footnotesize $x_P^+$}]{};
\node at (-2.4,-2.4) [label = right:{\footnotesize $f^{-1}(x^-)$}]{};
\node at (-1,1) [label = right:{\footnotesize $x^-$}]{};

\node at (-1.3,0.3) [circle, fill, inner sep=1.5 pt]{};
\node at (-2.2,-2.2) [circle, fill, inner sep=1.5 pt, blue]{};
\node at (-1/2,-1/2) [circle, fill, red, inner sep=1.5 pt]{};
\node at (-3.2/2,-3.2/2) [circle, fill, violet, inner sep=1.5 pt]{};

\node at (-2.2,1) [label=below:$\Sigma$]{};

\draw [-{Stealth[length=2mm]},red,decorate,decoration={snake, segment length=4mm, amplitude=0.7mm, post length=1mm}](-1.05,-1.05) -- (-1.5,1.5-2.1);
\draw [blue,decorate,decoration={snake, segment length=4mm, amplitude=0.7mm}](-1.9,-1.9) -- (-3,-0.8);
\draw [-{Stealth[length=2mm]},blue,decorate,decoration={snake, segment length=4mm, amplitude=0.7mm, post length=1mm}](-3,-0.8) -- (-2.5,-2.5+2.2);

\end{tikzpicture}
\caption{\footnotesize Case $x^-\geq 0$. The left-moving modes passing through $\Sigma$ are the left-moving modes emanating from the interval $[x_P^+,x^+]$ depicted in red. The right-moving modes passing through $\Sigma$ are the left-moving modes emanating from the interval $[f^{-1}(x^-),x_P^+]$ depicted in blue. Tracing over every modes on $\Sigma$ is therefore equivalent to tracing over left-moving modes on the interval $[f^{-1}(x^-),x^+]$. \label{fig:LR_modes_case1}}
\end{subfigure}
\quad \,
\begin{subfigure}[t]{0.48\linewidth}
\centering
\begin{tikzpicture}
\path
       +(3,3)  coordinate (IItopright)
       +(-3,3) coordinate (IItopleft)
       +(3,-3) coordinate (IIbotright)
       +(-3,-3) coordinate(IIbotleft)
      
       ;
\draw (IItopleft) --
          node[midway, above, sloped] {{\footnotesize$x^-=g(x^+)$}}
      (IItopright) --
          node[midway, above,sloped] {{\footnotesize $x^-=f(x^+)$}}
      (IIbotright) -- 
          node[midway, below, sloped] {}
      (IIbotleft) --
          node[midway, above , sloped] {{\footnotesize $x^-=f(x^+)$}}
      (IItopleft) -- cycle;

\draw (IItopleft) -- (0,0) -- (4.3/2,4.3/2);
\draw[blue] (4.3/2,4.3/2) -- (IItopright);

\draw[blue] (IIbotleft) -- (-2.5,-2.5);
\draw[red] (-2.5,-2.5) -- (-1.3/2,-1.3/2);
\draw (-1.3/2,-1.3/2) -- (0,0) -- (IIbotright);

\draw (IItopleft) -- (-3.5,3.5) ;
\draw (IItopright) -- (3.5,3.5) ;
\draw (IItopright) -- (3.5,3.5) ;
\draw (-3.5,3.3) -- (-3.5,3.5) -- node[midway, above, sloped] {$x^-$} (-3.3,3.5) ;
\draw (3.5,3.3) -- (3.5,3.5) -- node[midway, above, sloped] {$x^+$} (3.3,3.5) ;

\draw (-3,-2) to [bend right=12] (0.2,-1.5);
\draw[dotted] (0.2,-1.5) -- (-1.3/2,-1.3/2);
\draw[dotted] (-2.5,-2.5) -- (-3,-2);
\draw[dotted] (0.2,-1.5) -- (3,3-0.2-1.5);
\draw[dotted] (3,3-0.2-1.5) -- (4.3/2,4.3/2);

\node at (-0.45,-0.45) [label = left:{\footnotesize $x^+$}]{};
\node at (-2.7,-2.7) [label = right:{\footnotesize $x_P^+$}]{};
\node at (5.3/2,5.3/2) [label = left:{\footnotesize $f^{-1}(x^-)$}]{};
\node at (0.8,-0.8) [label = right:{\footnotesize $x^-$}]{};

\node at (0.2,-1.5) [circle, fill, inner sep=1.5 pt]{};
\node at (4.3/2,4.3/2) [circle, fill, inner sep=1.5 pt, blue]{};
\node at (-2.5,-2.5) [circle, fill, inner sep=1.5 pt, violet]{};
\node at (-1.3/2,-1.3/2) [circle, fill, inner sep=1.5 pt, red]{};

\node at (-1,-1.8) [label=below:$\Sigma$]{};

\draw [-{Stealth[length=2mm]},blue,decorate,decoration={snake, segment length=4mm, amplitude=0.7mm, post length=0.5mm}](3,2.15) -- (2.7,-2.7+5.15);
\draw [blue,decorate, decoration={snake, segment length=4mm, amplitude=0.7mm}](3,2.15) -- (-1,-1-0.85);

\draw [-{Stealth[length=2mm]},red,decorate,decoration={snake, segment length=4mm, amplitude=0.7mm}](-1,-1-0.85) -- (-1.3,1.3-2.85);
\draw [-{Stealth[length=2mm]},red](-2.25,-2.25) -- (-2.4,2.4-4.5);

\draw [blue,decorate,decoration={snake, segment length=4mm, amplitude=0.7mm}] (-2.75,-2.75) -- (-3,-2.5);
\draw [-{Stealth[length=2mm]},blue,decorate,decoration={snake, segment length=4mm, amplitude=0.7mm, post length=1.2mm}] (-3,-2.5) -- (-2.6,-2.6+0.5);

\end{tikzpicture}     
\caption{\footnotesize Case $x^-\leq 0$. The left-moving modes passing through $\Sigma$ are the left-moving modes emanating from (or which will cross) the interval $[x_P^+,x^+]$ depicted in red. The right-moving modes passing through $\Sigma$ are the left-moving modes emanating from (or which will cross) the interval $(-\infty,x_P^+] \cup [f^{-1}(x^-),+\infty)$ depicted in blue. Tracing over every modes on $\Sigma$ is therefore equivalent to tracing over left-moving modes on the interval $(-\infty,x^+]\cup [f^{-1}(x^-),+\infty)$ or on the complementary interval $[x^+,f^{-1}(x^-)]$. \label{fig:LR_modes_case2}}
\end{subfigure}
    \caption{\footnotesize Cauchy slice $\Sigma$ bounded by a point $(x_P^+,f(x_P^+))$ on the left boundary of the spacetime and a point $(x^+,x^-)$ in the bulk. Considering reflective boundary conditions at the timelike boundaries of the spacetime induces correlations between left and right-moving modes passing through $\Sigma$. \label{fig:LR_modes}}
\end{figure}
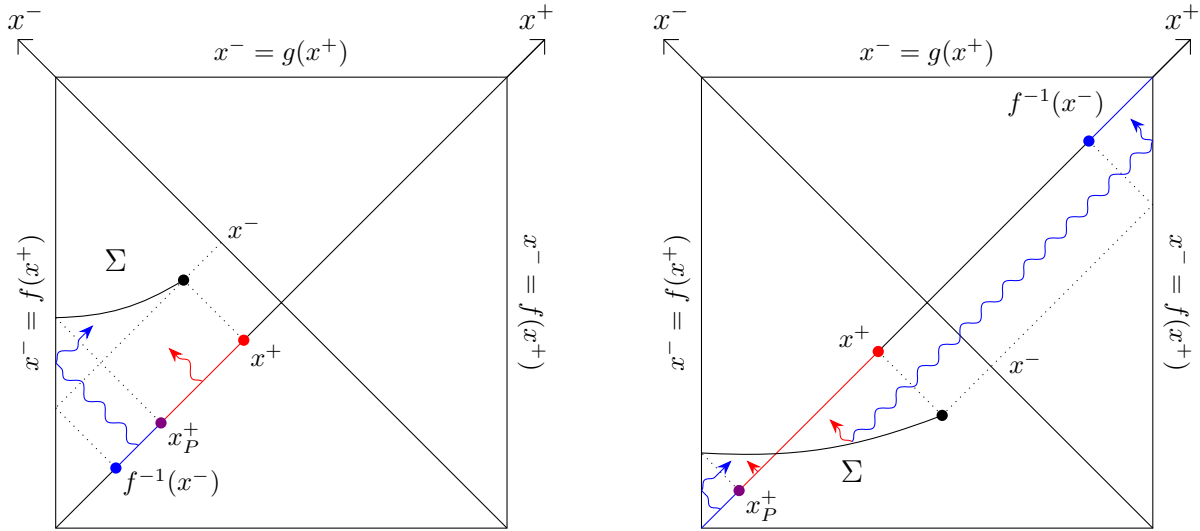

Similarly to the spatial boundaries, we parameterize $\mathcal{J}^{\pm}$ as a spacelike curve described by $x^- = g(x^+)$, with $g:I\rightarrow I$ a $C^1$ involution such that $g(x^-)x^-\geq 0$ and $-\infty < \dot g < 0$. From now on, we take the limit $L\rightarrow \infty$, which will be justified at the end of this section. In \Fig{fig:LR_modes},
the timelike and spacelike boundaries are depicted with examples of reflecting light rays, and we introduce the following notation to describe segments of the cosmological horizons:
\begin{align}
  \begin{aligned}
   H_-^- &= \{(x^+,x^-)|x^+=0,x^-\leq 0\}, \\   H_+^- &= \{(x^+,x^-)|x^-=0,x^+\leq 0\},
  \end{aligned}
  &&
  \begin{aligned}
   H_-^+ &= \{(x^+,x^-)|x^+=0,x^-\geq 0\}, \\   H_+^+ &= \{(x^+,x^-)|x^-=0,x^+\geq 0\},
  \end{aligned}
 \end{align}
so that $H_- = H_-^- \cup H_-^+$ is the horizons along $x^-$ and $H_+=H_+^-\cup H_+^+$ is the horizon along $x^+$.

From our construction, we can consider right and left-moving modes at $\mathcal{J}^-$ and note that all the right-moving modes pass through the half line $H_-^-$ while all the left-moving modes pass through the half line $H_+^-$. Because of the reflecting boundary conditions, all right-moving (left-moving) modes passing through $H_-^-$ ($H_+^-$) also cross the half line $H_+^+$ ($H_-^+$) as left-moving (right-moving) modes. In other words, all modes coming from $\mathcal{J}^-$ cross $H_-$ as right-movers and $H_+$ as left movers, such that we can express the quantum state of the fields as left-moving (right-moving) states at $H_-$ ($H_+$). From here we will always refer to the state of the fields as left-moving states on $H_+$. The condition that $f$ is an involution ensures that the following procedure is independent of the reference axis.

Now, consider a subsystem $\Sigma$ of some Cauchy slice $\Sigma_{\mathcal{M}}$, bounded by the left boundary of $\mathcal{M}$ at $(x_P^+,f(x_P^+)$) and some point $(x^+,x^-)$. As mentioned above, right and left-moving modes on $\Sigma$ are correlated and tracing over right and left-movers on $\Sigma$ is equivalent to tracing over left-movers on some interval of $H_+$. First, all left-moving modes on $\Sigma$ are the same modes as the left-moving modes on the interval $[x_P^+,x^+]$ of $H_+$. As for right-movers on $\Sigma$, the projection slightly varies depending on the position of the endpoint $(x^+,x^-)$:
\begin{itemize}
\item When $x^- \geq 0$, the right-moving modes on $\Sigma$ are the left-moving modes in the interval $[f^{-1}(x^-),x_P^+]$ of $H_+$.\footnote{We keep the notation $f^{-1}$ for clarity, although we imposed $f=f^{-1}$.} Therefore, tracing over every modes on $\Sigma$ is equivalent to tracing over left-moving modes on the interval $[f^{-1}(x^-),x^+]$ of $H_+$, see \Fig{fig:LR_modes_case1}. \item When $x^-\leq 0$, the right-moving modes on $\Sigma$ are the same as the left-moving modes in the subset $(-\infty,x_P^+] \cup [f^{-1}(x^-),\infty)$ of $H_+$. Therefore, tracing over every modes on $\Sigma$ is equivalent to tracing over left-moving modes on the interval $(-\infty,x^+] \cup [f^{-1}(x^-),\infty)$ of $H_+$, see \Fig{fig:LR_modes_case2}. Since the left-movers on $H_+$ should be in a pure state, this is equivalent of tracing over the complement of this subset, \ie $[x^+,f^{-1}(x^-)]$.
\end{itemize}
Assuming $|x^+-f^{-1}(x^-)|\ll L$, we can use \eq{eq:entropy_left} to get
\begin{equation}
    S(\Sigma) = \frac{c}{12}\ln\left[\frac{(x^+ -f^{-1}(x^-))^2}{\delta x^+~\delta f^{-1}(x^-)}\right],
\end{equation}
where $\delta x^+$ and $\delta f^{-1}(x^-)$ are the UV cutoffs at the endpoints of the interval of $H_+$. In other words, we have smeared over the regions $[x^+ -\delta x^+,x^+]$ and $[f^{-1}(x^-), f^{-1}(x^-)+ \delta f^{-1}(x^-)]$ on $H_+$ in the case $x_-\geq 0$, and similarly when $x_-\leq 0$. We would like to relate these quantities to the invariant cutoff $\delta^2$, \ie to the UV cutoff associated to the position of $(x^+,x^-)$. Since the left-moving modes on $[x^+ -\delta x^+,x^+]$ are the same as the left-moving modes on the same interval at $x^+$ fixed, we write $\delta x^+=\varepsilon_{\Le}(x^+,x^-)$. To relate the interval $[f^{-1}(x^-), f^{-1}(x^-)+ \delta f^{-1}(x^-)]$ on $H_+$ to an interval around $(x^+,x^-)$, we need to take into account the reflection on the left boundary. Considering the reflection of two light rays coming from the two endpoints of the interval, smearing over the left-moving modes on $[f^{-1}(x^-), f^{-1}(x^-)+ \delta f^{-1}(x^-)]$ on $H_+$ is equivalent to smearing over the right-moving modes on $[f(f^{-1}(x^-)), f(f^{-1}(x^-)+ \delta f^{-1}(x^-))]$ at fixed $x^+$ around the point $(x^+,x^-)$. At first order in $\delta f^{-1}(x^-)$, this interval is $[x^-,x^- +\delta f^{-1}(x^-) \dot f(f^{-1}(x^-))]$, from which we can identify the cutoff for the right movers at the point $(x^+,x^-)$ to be:
\begin{equation}
    \varepsilon_{\Ri}(x^+,x^-) = \delta f^{-1}(x^-)~\dot f(f^{-1}(x^-)).
\end{equation}
It is now possible to write $\delta x^+~\delta f^{-1}(x^-)$ in terms of an affine proper distance, and we conclude that
\begin{equation}
\label{eq:ent_bdy}
    S(\Sigma) = \frac{c}{6}\omega(x^+,x^-)+\frac{c}{12}\ln\left[\dot f(f^{-1}(x^-))\frac{(x^+ -f^{-1}(x^-))^2}{\delta^2}\right],
\end{equation}
where $\delta^2=\delta_{\Le}\delta_{\Ri}$ is the cutoff in inertial coordinates at the endpoint of $\Sigma$.

For this formula to be well defined, we need to motivate the assumption $|x^+-f^{-1}(x^-)|\ll L$. As $L\rightarrow \infty$, this is equivalent to imposing $-L\ll f^{-1}(x^-)\ll L$. For $f$ to break this condition, one would need to localize $x^-$ with an infinite precision as $L\rightarrow \infty$, which cannot be done in the presence of the UV cutoff $\delta$.

\subsection{Change of vacuum}
\label{sec:changevac}

As noted in \cite{Fiola:1994ir}, the entropy formula \eqref{eq:entropy_left} is only valid in the vacuum state defined with respect to the coordinates $x^{\pm}$, with the associated metric $\d s^2 = -e^{2\omega_x(x^+,x^-)}\d x^+\d x^-$, where we made the coordinate system explicit in the notation $\omega_x$. In general, the vacuum state may be defined with respect to an arbitrary set of lightcone coordinates $y^{\pm}(x^{\pm})$, in which the metric is $\d s^2 = -e^{2\omega_y(y^+,y^-)}\d y^+\d y^-$. Under this change of coordinates, the endpoints of $\Sigma$ become $(y^+(x_1^+),y^-(x_1^-))$ and $(y^+(x_2^+),y^-(x_2^-))$. By diffeomorphism invariance, the Cauchy slice defined by this interval in coordinates $y^{\pm}$ is physically identical to the Cauchy slice defined by the endpoints $(x_1^+,x_1^-)$ and $(x_2^+,x_2^-)$ in the original coordinates. For this reason, we call them both $\Sigma$, independently of the system of coordinates in which it is described.

To write the entropy contribution from the left-moving modes in a vacuum defined in coordinates $y^{\pm}$, one writes the size of the interval $[x_1^+,x_2^+]$ in $y^{\pm}$ coordinates \cite{Fiola:1994ir},
\begin{equation}
    S_y(\Sigma) = \frac{c}{12}\ln\left[\frac{(y^+(x_2^+)-y^+(x_1^+))^2}{\varepsilon_{\Le}(y(x_1))\varepsilon_{\Le}(y(x_2))}\right],
\end{equation}
where the index $y$ denotes the vacuum in which the entropy is computed. As we change the coordinate system, the cutoffs $\varepsilon_{\Le(\Ri)}$ must change according to:
\begin{equation}
    \varepsilon_{\Le(\Ri)}(y^+,y^-) = e^{-\omega_y(y^+,y^-)}~ \delta_{\Le(\Ri)}.
\end{equation}
Therefore, the entropy $S_y(\Sigma)$ of $\Sigma$ in the vacuum state defined in coordinates $y^{\pm}$ is given by \cite{Fiola:1994ir}:
\begin{align}
\begin{split}
\label{transf_entropy}
    S_{y}(\Sigma) &= \frac{c}{6}\left(\omega_y(y(x_1))+\omega_y(y(x_2))\right)+\frac{c}{12}\ln\left[\frac{(y^-(x^-_2)-y^-(x^-_1))^2(y^+(x^+_2)-y^+(x^+_1))^2}{\delta_1^2\delta_2^2}\right] \\
    &=\frac{c}{12}\left[2\left(\omega_x(x_1)+\omega_x(x_2)\right)+\ln\left[\left.\frac{\d x^+}{\d y^+}\right|_{x_1^+}\left.\frac{\d x^+}{\d y^+}\right|_{x_2^+}\left.\frac{\d x^-}{\d y^-}\right|_{x_1^-}\left.\frac{\d x^-}{\d y^-}\right|_{x_2^-}\right] \right.\\
    &+\left.\ln\left[\frac{(y^+(x^+_2)-y^+(x^+_1))^2(y^-(x^-_2)-y^-(x^-_1))^2}{\delta_1^2\delta_2^2}\right]\right],
\end{split}
\end{align}
where we have used \eqref{eq:transfo_conf_factor}. The formulas above should be applied to spacetimes that have infinite extends in both $x^+$ and $x^-$ (or $y^+$ and $y^-$) directions. In particular it does not take into account any possible boundaries. 

A similar transformation law can be derived in the presence of spatial boundaries. First, supposing that the boundary in coordinates $x^{\pm}$ follows a trajectory $x^-=f_x(x^+)$, the spatial boundary in coordinates $y^{\pm}$ is described by
\begin{equation}
    y^- = f_y(y^+) =y^-(f_x(x^+)).
\end{equation}
From this, we can compute the transformation law of the term $\dot f_x(f_x^{-1}(x^-))$:
\begin{equation}
    \dot f_x(f_x^{-1}(x^-)) \rightarrow \dot f_y(f_y^{-1}(y^-)) = \dot f_x(f_x^{-1}(x^-))\frac{\frac{\d y^-}{\d x^-}(x^-)}{\frac{\d y^+}{\d x^+}(f_x^{-1}(x^-))}.
\end{equation}
Hence, the entropy of a Cauchy slice $\Sigma$ bounded by a point $(x^+,x^-)$ and the boundary \linebreak $x^-=f_x(x^+)$, on a curved background with the metric $\d s^2 =-e^{2\omega_x}\d x^+\d x^-$ and in a vacuum state defined in coordinates $y^{\pm}(x^{\pm})$, is given by:
\begin{align}
\begin{split}
\label{eq:transf_bdy}
    S_{y}(\Sigma) &= \frac{c}{6}\omega_x(x^+,x^-) + \frac{c}{12}\ln\left[\dot f_x(f_x^{-1}(x^-))\frac{\left(y^+(x^+) -y^+(f_x^{-1}(x^-))\right)^2}{\delta^2}\right]\\
    &-\frac{c}{12}\ln\left[\frac{\d y^+}{\d x^+}(x^+)\frac{\d y^+}{\d x^+}\left(f_x^{-1}(x^-)\right)\right].
\end{split}
\end{align}

\section{Light-sheet, expansion and energy conditions}
\label{sec:hydro_regime}

In this Appendix, we recall the formalism describing the classical Bousso bound in the hydrodynamic approximation, following \cite{Flanagan:1999jp,Bousso:2003kb}. We will first consider an $(n+1)$-dimensional framework, and then carry out the dimensional reduction yielding the classical Bousso bound in JT gravity, that is considered in \Sect{sec:lightsheets}. 

We consider a light-sheet $L(B-B')$, emanating and ending from the codimension $2$ surfaces $B$ and $B'$, respectively. We call $\lambda$ the affine parameter along each null geodesic generating $L(B-B')$, normalized such that $\lambda=0$ on $B$ and $\lambda=1$ on $B'$. Since $\d/\d\lambda$ is null and orthogonal to the light-sheet by construction, we define the future directed vector normal to $L(B-B')$ by $k^{M} = \pm \left(\frac{\d}{\d\lambda}\right)^{M}$, with a ``$+$'' sign for future-directed light-sheets and a ``$-$'' sign for past-directed light-sheets. For any choice of affine parameter, the normal vector must satisfy the geodesic equation
\begin{equation}\label{eq:k_geo_eq}
    k^{M}\hat\nabla_{M}k^{N} = 0,
\end{equation}
with $\hat \nabla_M$ the covariant derivative compatible with the metric $\hat g_{MN}$ of $\hat{\mathcal{M}}$. One can introduce the induced metric $h_{ab}^{(n)}$ and the extrinsic curvature $K_{ab}^{(n)}$ of the $n$-dimensional hypersurface $L(B-B')$ of normal vector $k^M$, respectively by:
\begin{eqnarray}
    h_{ab}^{(n)} &=&\frac{\partial X^M}{\partial y^{a}}\frac{\partial X^N}{\partial y^{b}}\hat{g}_{MN},\\
    K^{(n)}_{ab} &=& \frac{\partial X^{M}}{\partial y^{a}} \frac{\partial X^{N}}{\partial y^{b}}\hat \nabla_{M} k_{N},
\end{eqnarray}
where $\{X^{M},~M=0,...,n\}$ are a set of coordinates on $\hat{\mathcal{M}}$ and $\{y^{a},~a=0,...,n-1\}$ a set of coordinates on $L(B-B')$. With these definitions, the expansion parameter of the congruence is defined as the trace $h_{(n)}^{ab}K_{ab}^{(n)}$ of the extrinsic curvature tensor\cite{Wald:1984rg},
\begin{equation}
\label{eq:defcurv}
    \theta^{(n)} = \hat\nabla_{M} k^{M},
\end{equation}
where the superscript $(n)$ indicates that $\theta^{(n)}$ is the expansion parameter of an $n$-dimensional null congruence. By definition of a light-sheet, $\theta^{(n)}$ is non-positive everywhere on it, \linebreak $\theta^{(n)}(\lambda)\leq 0, \forall \lambda\in [0,1]$.

The hydrodynamic approximation relies upon a local entropy current $\hat s^{M}$. The $n$-dimensional entropy flux $s^{(n)}$ on the light-sheet is given by
\begin{equation}\label{eq:entropy_flux_n}
s^{(n)} =-k^M \hat s_M.
\end{equation}
The integral of $s^{(n)}$ over $L(B-B')$ gives the total matter entropy passing through $L(B-B')$. Considering a coordinate system $y=(y^1,...,y^{n-1})$ on $B$, each null geodesic generating the light-sheet is thus labelled by $y$, and $(y^1,...,y^{n-1},\lambda)$ provides a coordinate system on $L$. We denote by $h(y,\lambda)$ the determinant of the induced metric on any fixed $\lambda$ cross-section of $L(B'-B)$. It is related to the determinant $h(y,0)$ of the induced metric on $B$ by
\begin{equation}
h(y,\lambda)=\left[\mathcal{A}^{(n)}(y,\lambda)\right]^2~h(y,0),
\end{equation}
where $\mathcal{A}^{(n)}(y,\lambda)$ is the area decrease factor, defined in terms of the expansion scalar $\theta(\lambda)$ of a given generator of $L$ by:
\begin{equation}\label{eq:area_decrease_factor}
\mathcal{A}^{(n)}\equiv\exp\left[\int_0^{\lambda}\d\tilde{\lambda}~\theta^{(n)}(\tilde{\lambda})\right].
\end{equation}
Equivalently, $\theta^{(n)}$ is the logarithmic derivative of $\mathcal{A}^{(n)}$, $\theta^{(n)}(\lambda)=\d\ln\mathcal{A}^{(n)}/\d\lambda$. By definition of a light-sheet, $\theta^{(n)}(\lambda)\leq 0$, or equivalently $\d \mathcal{A}^{(n)}/ \d\lambda\leq 0$, $\forall\lambda$.

With these notations, the total entropy flux passing through the light-sheet $L(B-B')$ reads:
\begin{eqnarray}\label{eq:4D_entropy_flux}
    \int_{L(B-B')} s^{(n)} &=& \int_B \d^{n-1}y \int_0^1 \d\lambda~\sqrt{h(y,\lambda)}~s^{(n)}(y,\lambda)\nonumber\\
    &=&\int_B \d^{n-1}y \sqrt{h(y,0)}\int_0^1 \d\lambda~\mathcal{A}^{(n)}(y,\lambda)~s^{(n)}(y,\lambda).
\end{eqnarray}
With this formal definition, the Bousso bound states that the coarse-grained entropy \eqref{eq:4D_entropy_flux} passing through $L(B-B')$ satisfies the inequality
\begin{equation}
\label{eq:GCEB}
\int_{L(B-B')} s^{(n)} \leq \frac{1}{4\hat{G}}(A(B)-A(B')).
\end{equation}
As shown in \cite{Flanagan:1999jp}, a sufficient condition to prove the generalized covariant entropy bound \eqref{eq:GCEB} is to derive the inequality
\begin{equation}\label{eq:sufficient_cond_GCEB}
\int_0^1 \d\lambda~\mathcal{A}^{(n)}(y,\lambda)~s^{(n)}(y,\lambda)\leq \frac{1}{4\hat{G}}(\mathcal{A}^{(n)}(0)-\mathcal{A}^{(n)}(1)),
\end{equation}
for each null generator of $L$, \ie at fixed $y$. In the following we will therefore drop the $y$ dependence of $\mathcal{A}^{(n)}$ and $s^{(n)}$. Note that in JT gravity, the area decrease factor along the light ray reduces to
\begin{equation}\label{eq:area_decrease_factor_2d}
    \mathcal{A}(\lambda) = \frac{\Phi(\lambda)}{\Phi(0)}.
\end{equation}
This is obtained by inserting the relation \eqref{eq:thJT} into the general definition \eqref{eq:area_decrease_factor} of the area decrease factor.

The hydrodynamic approximation cannot be fundamental, since the entropy is a non-local notion. It is therefore only valid in certain regimes and over certain scales. In particular, it is a good approximation only if the size of the system whose entropy is computed is much larger than some microscopic length scale. To ensure its validity, two conditions can be imposed, that have been shown to be sufficient to prove the Bousso bound \cite{Flanagan:1999jp, Bousso:2003kb, Strominger:2003br}:
\begin{enumerate}
    \item The ``\emph{gradient of entropy density}'' condition:
    \begin{equation}
    \label{eq:cond1}
        \left| k^M k^N\hat\nabla_M \hat s_N\right| \leq 2\pi k^M k^N \braket{\hat T_{MN}},
    \end{equation}
    where $\braket{\hat T_{MN}}$ is the stress-energy tensor in $\hat{\mathcal{M}}$. Using the property \eqref{eq:k_geo_eq} of the vector $k^M$, this inequality can be rewritten in terms of the entropy flux $s^{(n)} =-k^M \hat s_M$:
    \begin{equation}
    \label{eq:cond1_bis}
        \left| (s^{(n)}(\lambda))' \right| \leq 2\pi k^M k^N \braket{\hat T_{MN}},
    \end{equation}
    where $'\equiv \d/\d\lambda=k^{M}\hat\nabla_{M}$.
    \item The ``\emph{isolated system}'' condition:
    \begin{equation}
    \label{eq:cond2}
        s^{(n)}(0) \leq -\frac{\theta^{(n)}(0)}{4 \hat{G}}.
    \end{equation}
\end{enumerate}
The first condition is necessary for the whole phenomenological description of entropy in terms of entropy densities to be valid. It requires that the gradient of the entropy density is bounded by the energy density. This ensures that $\hat s^M$ only varies significantly over length scales greater than the largest wavelength of any of the modes that contribute to the entropy \cite{Flanagan:1999jp, Bousso:2003kb}. Below this scale, there is no meaning for a variation of the entropy density. Notably, this condition implies the NEC
\begin{equation}
\label{eq:neq}
    k^M k^N \braket{\hat T_{MN}} \geq 0.
\end{equation}
The second condition was first stated in \cite{Bousso:2003kb} as $s^{(n)}(0)=0$. They argued that it does not add any notable restriction to the system, since it should always be possible to modify slightly the entropy current around the initial surface without breaking the first condition nor changing the total integrated entropy by a non-negligible amount. The modified condition \eqref{eq:cond2}, corresponding to imposing that the bound is valid at the beginning of the light-sheet, was used in \cite{Strominger:2003br} to prove the Bousso bound in the Callan-Giddings-Harvey-Strominger (CGHS) model \cite{Callan:1992rs} with similar motivations. Notably, this modification implicitly requires the light-sheet to be initially non-expanding, \ie $\theta^{(n)}(0)\leq 0$. Imposing these ``isolated system'' conditions is equivalent to requiring that modes carrying a non-negligible contribution to the entropy should be fully contained on the light-sheet, and not spanned beyond the initial surface \cite{Bousso:2003kb}.

On the other hand, the Raychaudhuri equation for $n>1$ determines the evolution of the expansion parameter $\theta^{(n)}$ along the congruence:
\begin{equation}
\frac{\d\theta^{(n)}}{\d\lambda}=-\frac{1}{n-1}(\theta^{(n)})^2-\hat\sigma_{MN}\hat\sigma^{MN}+\hat\omega_{MN}\hat\omega^{MN}-\hat R_{MN} k^{M}k^{N},
\end{equation}
where $\hat \sigma_{MN}$ and $\hat \omega_{MN}$ are the shear and twist tensors respectively. For a congruence hypersurface orthogonal, $\hat\omega_{MN}=0$. Assuming in addition that the NEC $\braket{\hat T_{MN}}k^{M}k^{N}\geq 0$ is satisfied for any null vector $k^{N}$, then the Einstein equations imply that $\hat R_{MN}k^{M} k^{N}\geq 0$, and the Raychaudhuri equation leads to the Classical Focussing Theorem:
\begin{equation}
\label{eq:clfocus}
\frac{\d\theta^{(n)}}{\d\lambda}\leq 0.
\end{equation}
Physically, this tells us that the light rays are focused during the evolution of the congruence. Since the two conditions \eqref{eq:cond1} and \eqref{eq:cond2} require the NEC \eqref{eq:neq} and initial non-expansion, this motivates the fact that the Bousso bound does not apply to null congruences of positive expansion. Moreover, supposing the NEC, we can reduce the definition of a light-sheet to a null congruence that is initially non-expanding.

The Bousso bound and the entropy conditions in the JT gravity setup are obtained from their $(n+1)$-dimensional counterparts using the spherical reduction \eqref{eq:sphred}. The two-dimensional entropy flux $s$ is related to the $n$-dimensional entropy flux $s^{(n)}$ by
\begin{equation}\label{eq:2d_entropy}
s(\lambda)=S_{n-1}(l_n)\Phi(\lambda)~s^{(n)}(\lambda),
\end{equation}
where $S_{n-1}(l_n)=2\pi^{n/2}~l_n^{n-1}/\Gamma(n/2)$ is the surface area of the $(n-1)$-sphere of radius $l_n$. The total entropy flux \eqref{eq:4D_entropy_flux} passing through the light-sheet then rewrites as
\begin{align}
\int_{L(B-B')} s^{(n)} &= l_n^{n-1}\Phi(0) \int \d^{n-1}\Omega \int_0^1 \d\lambda~s^{(n)}(\lambda)\frac{\Phi(\lambda)}{\Phi(0)}\\
&= S_{n-1}(l_n)\int_0^1 \d\lambda~s^{(n)}(\lambda)\Phi(\lambda)\\
&= \int_0^1\d\lambda~s(\lambda).
\end{align} 
Using \eq{eq:area_decrease_factor_2d}, the dimensional reduction of the $(n+1)$-dimensional Bousso bound \eqref{eq:GCEB} gives the Bousso bound in two-dimensional JT gravity given in \eqref{eq:2D_GCEB}\footnote{Equivalently, we could have derived this inequality by inserting the relations \eqref{eq:area_decrease_factor_2d}, \eqref{eq:2d_entropy} and \eqref{eq:2d_Newton_cste} into the sufficient condition \eqref{eq:sufficient_cond_GCEB} of \cite{Flanagan:1999jp}, immediately yielding the Bousso bound in JT gravity.}. Similarly, the two entropy conditions \eqref{eq:cond1JT} and \eqref{eq:cond2JT} can be obtained from their $(n+1)$-dimensional counterparts \eqref{eq:cond1} and \eqref{eq:cond2} using \eqref{eq:2d_entropy} and    
\begin{equation}
T_{\mu\nu} = S_{n-1}(l_n)
\Phi~\hat T_{\mu\nu},
\end{equation} 
where $T_{\mu\nu}$ and $\hat T_{\mu\nu}$ are the stress-energy tensors in $(1+1)$ and $(n+1)$ dimensions respectively.

\bibliographystyle{jhep}
\bibliography{bib}

\end{document}